\documentclass[ aps, twocolumn, nofootinbib, showkeys, notitlepage, superscriptaddress]{revtex4-2}

\usepackage{natbib}
\usepackage{lipsum}
\usepackage{graphicx}
\usepackage{dcolumn}
\usepackage{bm}
\usepackage{amsmath}
\usepackage{amssymb}
\usepackage{float}
\usepackage{multirow}
\usepackage{slashed}
\usepackage{xcolor}
\usepackage{physics}
\usepackage{multirow}
\usepackage{gensymb}
\usepackage[colorlinks=true, pdfstartview=FitV, bookmarks=true, bookmarksnumbered=true, breaklinks]{hyperref}
\usepackage{mathtools,braket}
\usepackage{lipsum}  
\usepackage{color}
\definecolor{blue}{rgb}{0.0, 0.0, 1.0}
\definecolor{red}{rgb}{1.0, 0.0, 0.0}
\definecolor{royalblue}{rgb}{0.0, 0.14, 0.4}
\hypersetup{linkcolor=blue, citecolor=blue, urlcolor=blue}

\def\orcid#1{\kern .08em\href{https://orcid.org/#1}{\includegraphics[keepaspectratio,width=0.7em]{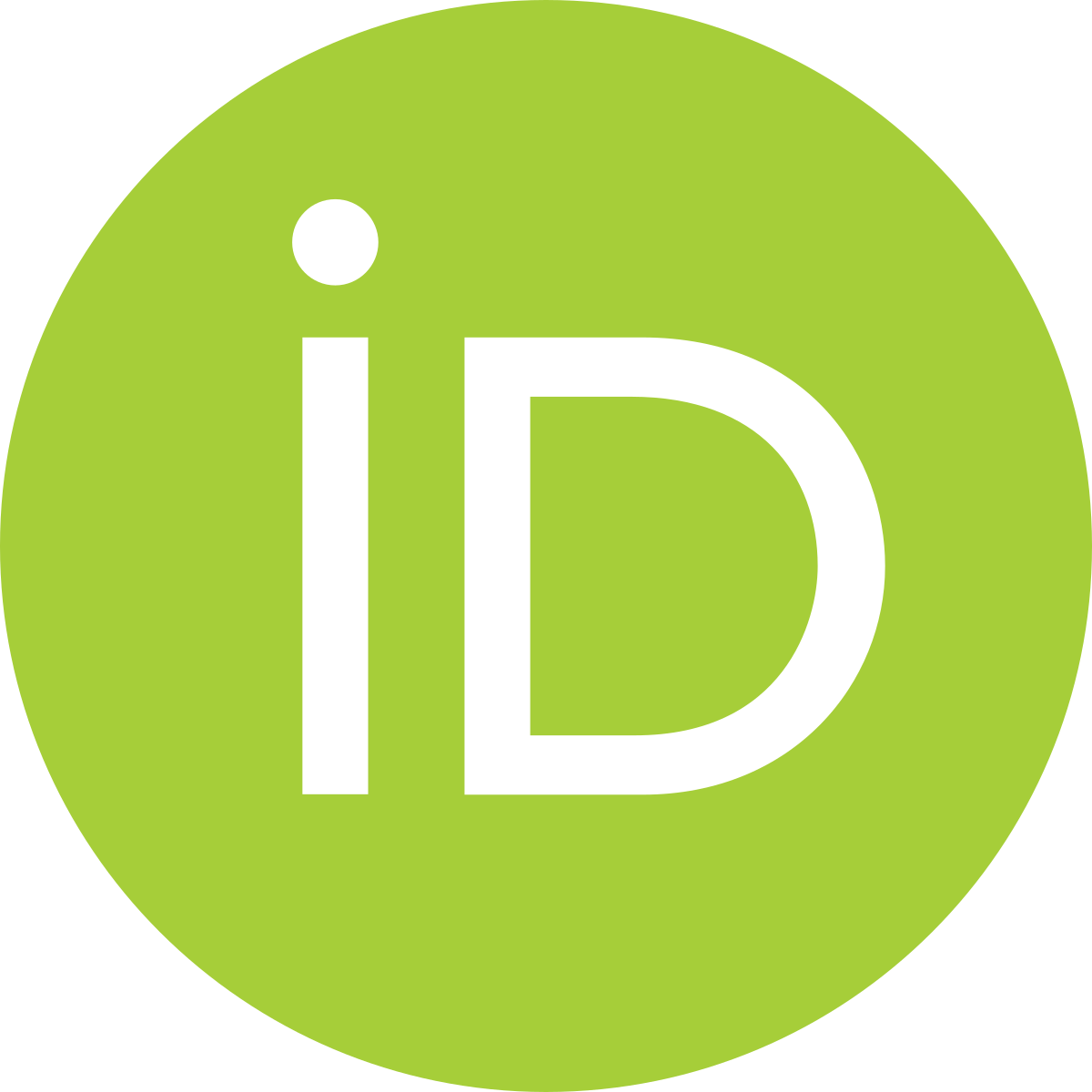}}}

\begin{document}

\title{Non-Hermitian Schwinger effect in an imaginary magnetic field}

\author{Ahmad Jafar Arifi\orcid{0000-0002-9530-8993}}
\email{aj.arifi01@gmail.com}
\affiliation{Advanced Science Research Center, Japan Atomic Energy Agency, Tokai 319-1195, Japan}
\affiliation{Research Center for Nuclear Physics, The University of Osaka, Ibaraki, Osaka 567-0047, Japan}

\author{Kei Suzuki\orcid{0000-0002-8746-4064}}
\email{k.suzuki.2010@th.phys.titech.ac.jp}
\affiliation{Advanced Science Research Center, Japan Atomic Energy Agency, Tokai 319-1195, Japan}

\date{\today}

\begin{abstract}
We investigate a non-Hermitian Schwinger effect by analyzing a one-loop effective Lagrangian of scalar and Dirac quantum electrodynamics (QED) in the presence of a real electric field and a parallel imaginary magnetic field.
We show that an imaginary magnetic field produces an additional pole family in the integrand of the proper-time representation.  
Unlike a real magnetic field, an imaginary magnetic field can generate a nonzero imaginary part of the effective Lagrangian even in the absence of an electric field.
The interplay between the two fields leads to nontrivial phenomena, such as a cancellation mechanism between the contributions of two fields.
We also point out that when the ratio of two field strengths is rational, some poles among the two families coincide, and a careful treatment of double poles is required.
By further examining the complex Landau-level structure of the electric-pole contribution, we find that the field dependence of the vacuum decay rate from each Landau level exhibits a sinusoidal-like behavior, and their sum determines the overall field dependence.
Our results clarify the analytic structure of the QED effective Lagrangian in imaginary magnetic backgrounds and provide a basis for investigating its relation to non-Hermitian phenomena in the QED vacuum.
\end{abstract}

\maketitle

\section{Introduction}

The \emph{Schwinger effect}~\cite{Sauter:1931zz, Heisenberg:1936nmg, Weisskopf:1936, Schwinger:1951nm}, a landmark prediction of quantum electrodynamics (QED), refers to the vacuum instability or the particle-antiparticle pair creation from the vacuum by a strong electric field (see Refs.~\cite{Dunne:2004nc, Ruffini:2009hg, Gelis:2015kya, Copinger:2020nyx,Fedotov:2022ely, Hattori:2023egw, Taya:2026llb} for reviews).
This phenomenon is a cornerstone of \emph{strong-field physics} and continues to drive research in a wide range of fields, including high-intensity laser physics~\cite{Yakimenko:2018kih, Danson:2019abc}.
Despite its long history, the Schwinger mechanism remains a highly active field of study rich with open questions.
Non-Hermitian environments occur in nature and can be controllably realized in experiments, but how the Schwinger effect is modified in such environments remains unknown.

Non-Hermitian quantum mechanics has emerged as a powerful framework for describing open systems, dissipative processes, and instability phenomena~\cite{Bender:2007nj,El-Ganainy:2018ksn,Bergholtz:2019deh,Ashida:2020dkc,Ding:2022juv}.
A typical example is the Hatano-Nelson model~\cite{Hatano:1996ntg,Hatano:1997}, where an imaginary vector potential was introduced through an asymmetric hopping term.
Systems with an imaginary vector potential were experimentally found~\cite{Weidemann2020,Liu2022}.
More generally, the spatial curl and temporal derivative of an imaginary vector potential can be formally interpreted as \emph{imaginary magnetic and electric fields}, respectively (for phenomenological discussion on imaginary fields, see Refs.~\cite{Ozawa:2023oqc,WANG20264112,Montag:2026cqh,Arifi:2026ume,Alminas:2026vhl}).
Such situations correspond to inhomogeneous and dynamical generalizations of the Hatano-Nelson model and are experimentally accessible.
Therefore, such developments raise a fundamental question: how does an imaginary electromagnetic field influence quantum phenomena in relativistic quantum field theory, particularly \emph{vacuum instability}?

To understand this influence, it is instructive to recall how a real magnetic field affects the Schwinger effect~\cite{Nikishov:1969tt,Bunkin:1969if,Marinov:1972nx,Daugherty:1976mg,Claudson:1980yz,Suganuma:1990nn,Ivanov:1992np,Suganuma:1991ha,Valluri:1993yi,Soldati:1998vq,Gavrilov:1998hw,Lin:1998rn,Hounkonnou:2000im,Cho:2000ei,Bae:2000pw,Jentschura:2001qr,Nikishov:2001fd,Kruglov:2001dp,Kruglov:2001cx,Kim:2003qp,Kim:2008yt,Tanji:2008ku,Iwazaki:2009bg,Fukushima:2010vw,Tanji:2011di,Warringa:2012bq,Bolognesi:2012gr,Sato:2013pxa,Hashimoto:2014yya,Ozaki:2015yja,Ghodrati:2015rta,Copinger:2016llk,Bavarsad:2017oyv,Copinger:2018ftr,Korwar:2018euc,Sheng:2018jwf,Karabali:2019oxq,Karabali:2019ucc,Zhu:2019igg,Hattori:2020guh,Taya:2020bcd,Bhattacharya:2020sjr,Copinger:2020feb,Ali:2021jch,Kaushal:2022las,Copinger:2022gfz,Copinger:2022bwl,Yu:2023cic,Kaushal:2024cbk,Dunne:2025cyo,Gupta:2025srv,Tahery:2026ehm}. 
While a real magnetic field does not independently produce particles from the vacuum, 
it can modify the vacuum decay rate induced by an electric field. 
In the proper-time representation of the effective Lagrangian, a magnetic field introduces hyperbolic structures, 
such as $1/\tanh(eBs)$ for Dirac QED and $1/\sinh(eBs)$ for scalar QED, which reflect Landau-level quantization and the corresponding modification of the vacuum spectrum. 

In this work, we investigate the Schwinger effect in scalar and Dirac QED in the presence of a real electric field and a parallel imaginary magnetic field.
This background is non-Hermitian
and modifies the proper-time structure of the effective Lagrangian.
As a result, we obtain intuitive formulas of the imaginary part in the manner of Schwinger's original formula.
Using these formulas, we demonstrate that (i) the emergence of a nonzero imaginary part of the effective Lagrangian even at vanishing electric field, which may be referred to as \emph{imaginary magnetic Schwinger effect}; (ii) the nontrivial interplay between the two fields, such as a cancellation mechanism of singularities; and (iii) the role of the complex Landau-level structure induced by an imaginary magnetic field.
These findings clarify the analytic structure of the effective Lagrangian in an imaginary magnetic background and its implications for vacuum instability.

The remainder of this paper is organized as follows.
In Sec.~\ref{sec:schwinger-complex}, after the standard Schwinger effect is reviewed, we derive the effective Lagrangians in an imaginary magnetic field, their imaginary parts, and complex Landau level representation. Section~\ref{sec:results} presents the numerical results and their physical interpretation. Section~\ref{sec:conclusion} summarizes our findings and outlook.
Appendix~\ref{app:schwinger} provides an additional review of the standard Schwinger effect, while Appendices~\ref{app:landau} and \ref{app:landau_imag} provide further discussions of the real and complex Landau levels, respectively. 
Finally, Appendix~\ref{sec:Hatano-Nelson} discusses the realization of an effective imaginary magnetic field in a Hatano-Nelson lattice.

\section{Schwinger effect in \\ imaginary magnetic fields}
\label{sec:schwinger-complex}

In this section, we study the one-loop effective Lagrangians for scalar and Dirac QED in an imaginary magnetic field.
These are derived by analytically continuing the proper-time representation known for a real magnetic field.
Using the derived effective Lagrangians, we discuss the structure of the imaginary part.
Before doing that, we first briefly review the standard Schwinger effect and its modification in the presence of a real magnetic field, where we assume static and uniform electromagnetic fields.

\begin{figure*}[t]
    \centering
    \includegraphics[width=2\columnwidth]{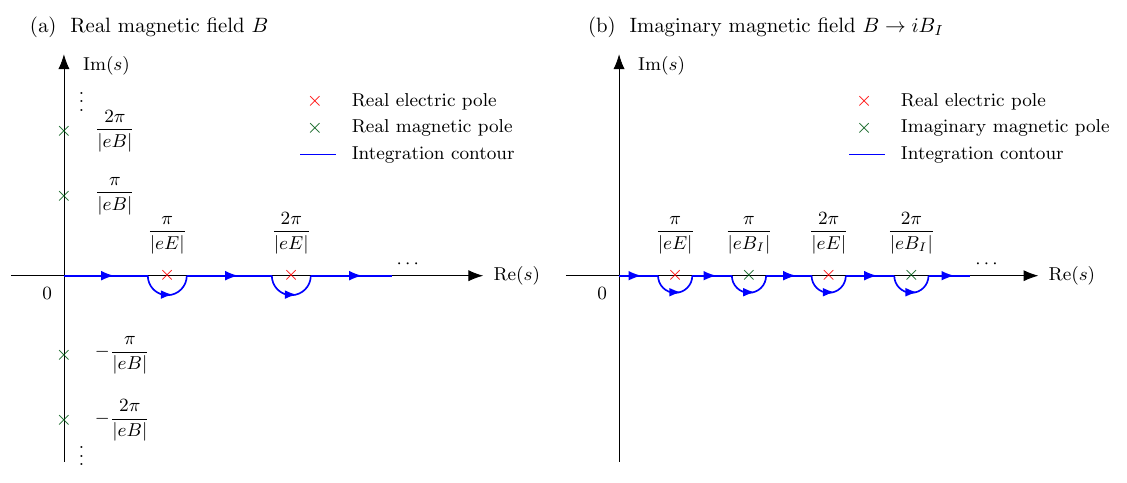} 
    \caption{Contour prescription for the proper-time integral. (a) For a real magnetic field, 
    the electric $E$ poles lie on the real axis and contribute to the imaginary part of effective Lagrangian through the 
    deformed integration contour, while the magnetic $B$ poles remain on the imaginary axis. 
    (b) After the analytic continuation $B\to iB_I$, the magnetic $B_I$ poles move onto 
    the real axis and also contribute to the imaginary part.}
    \label{fig:contour}
\end{figure*}

\subsection{Standard Schwinger effect}
\label{sec:schwinger}

We start with the case of parallel electric and magnetic fields. 
In the natural units ($\hbar=c=1$), the effective Lagrangians for 
scalar and Dirac QED are given by~\cite{Weisskopf:1936}
\begin{align}
&\mathcal{L}_{\text{scalar}}(E, B) = \frac{1}{16\pi^2} \int_0^\infty \frac{ds}{s^3} e^{-m^2 s} \nonumber\\
&\times\left[ \frac{(eEs)(eBs)}{\sin(eEs - i\epsilon)\sinh(eBs)} - 1 - \frac{e^2(E^2 - B^2)s^2}{6} \right], \label{eq:L_scalar_EB}
\end{align}
and~\cite{Heisenberg:1936nmg}
\begin{align}
&\mathcal{L}_{\text{Dirac}}(E, B) = -\frac{1}{8\pi^2} \int_0^\infty \frac{ds}{s^3} e^{-m^2 s} \nonumber\\
&\times\left[ \frac{(eEs)(eBs)}{\tan(eEs - i\epsilon)\tanh(eBs)} - 1 + \frac{e^2(E^2 - B^2)s^2}{3} \right], \label{eq:L_Dirac_EB}
\end{align}
respectively, where $m$ is the particle mass, $e>0$ is the elementary charge, and $s$
denotes the Schwinger proper-time parameter. 
The second and third terms represent the vacuum subtraction and
charge-renormalization terms, respectively.
The electric-field factors
$\sin(eEs-i\epsilon)$ and $\tan(eEs-i\epsilon)$ generate a sequence of
poles along the positive real proper-time axis at~\cite{Heisenberg:1936nmg}%
\footnote{In our convention, the electric poles lie on the positive real
$s$ axis, whereas the magnetic poles lie on the imaginary axis.
In a different convention, such as under the transformation $s=-i\tau$, the
locations of the electric and magnetic poles are interchanged, together
with the corresponding contour-integration prescription.}
\begin{equation}
    s_n=\frac{n\pi}{|eE|}, \qquad n=1,2,3,\ldots ,
\end{equation}
where $n$ labels the poles of the proper-time integrand. The
$-i\epsilon$ prescription specifies how the integration contour avoids
these poles, as illustrated in Fig.~\ref{fig:contour}(a). The resulting
pole contributions generate the imaginary part of the effective Lagrangian.
Since the vacuum persistence probability is $\left|e^{iVT\mathcal{L}}\right|^2=e^{-2VT \mathrm{Im}\mathcal{L}}$ in a four volume $VT$, the vacuum decay rate (per unit four volume) $\Gamma \equiv -\frac{1}{VT} \ln\left|e^{iVT\mathcal{L}}\right|^2$ is $\Gamma = 2\,\mathrm{Im}\,\mathcal{L}$~\cite{Schwinger:1951nm}.
In contrast, for a real magnetic field, the magnetic poles lie away from
the integration contour and therefore do not contribute directly to the
imaginary part. The magnetic field instead modifies the residues at the
electric poles through the hyperbolic factors in
Eqs.~\eqref{eq:L_scalar_EB} and \eqref{eq:L_Dirac_EB}, thereby modifying
the vacuum decay rate.
The detailed discussions of the imaginary parts are provided in Appendix~\ref{app:schwinger}.

\subsection{Imaginary magnetic field}
\label{sec:imag}

By performing the analytic continuation of $B \to iB_I$,\footnote{In this work, we focus on pure imaginary magnetic fields.
The extension to a complex magnetic field, $B\to B_R+iB_I$, is also possible.} the hyperbolic factors in Eqs.~\eqref{eq:L_scalar_EB} and \eqref{eq:L_Dirac_EB} become
\begin{align}
\sinh(ieB_I s) &= i\sin(eB_I s),\label{eq:sinh_to_sin} \\
\tanh(ieB_I s) &= i\tan(eB_I s). \label{eq:tanh_to_tan}
\end{align}
One can immediately see that the poles originating from the imaginary magnetic field, 
which lie on the imaginary axis for real magnetic field, are shifted onto 
the real axis after the analytic continuation as depicted in Fig.~\ref{fig:contour}(b). 
Note that the imaginary units introduced by the analytic continuation in the numerators and denominators in the first terms of Eqs.~\eqref{eq:L_scalar_EB} and \eqref{eq:L_Dirac_EB} cancel each other.

\subsubsection{Scalar QED}

In scalar QED, the one-loop effective Lagrangian in an imaginary magnetic field can be written as
\begin{align}
&\mathcal{L}_{\text{scalar}}(E,B_I)
= \frac{1}{16\pi^2}\int_0^\infty \frac{ds}{s^3}e^{-m^2s}
\nonumber\\
&\times
\left[
\frac{e^2EB_I s^2}
{\sin(eEs-i\epsilon)\sin(eB_Is-i\epsilon)}
-1 -\frac{e^2(E^2+B_I^2)s^2}{6}
\right].
\label{eq:L_scalar_EBI}
\end{align}
This form can be regarded as an analytic continuation from the real-magnetic-field case, Eq.~\eqref{eq:L_scalar_EB}.
Thus, the effective Lagrangian is characterized by the product of two cosecant factors.\footnote{A related system is the Euclidean self-dual background, where the Euclidean field strength tensor is constrained by the self-duality condition~\cite{Belavin:1975fg} $F_{\mu\nu}=\tilde{F}_{\mu\nu}$.
In Minkowski spacetime, it can be interpreted as a parallel configuration, such as the magnetic case $|eB|=|eE_I|$ and the electric case $|eE|=|eB_I|$.
In this case, the effective Lagrangians~\cite{Dunne:2001pp,Dunne:2002qf,Dunne:2002qg,Dunne:2002ta,Dunne:2003tr} have $1/\sinh^2(eBs)$ and $1/\sin^2(eEs)$, respectively.
Therefore, the Lagrangian~\eqref{eq:L_scalar_EBI} with $1/\sin(eEs)\sin(eB_Is)$ can be regarded as a generalized form of the latter.}

Here, we emphasize that although $\mathcal{L}(E,B_I)$ can be derived by the analytic continuation $B\to iB_I$, in general, its imaginary part $\mathrm{Im}\,\mathcal{L}(E,B_I)$ cannot be obtained by simply substituting $B\to iB_I$ into $\mathrm{Im}\,\mathcal{L}(E,B)$.
This is because, under this continuation, the magnetic-field poles move onto the proper-time integration contour.
Therefore, to correctly calculate the imaginary part, we need an appropriate pole prescription, which corresponds to introducing the $-i\epsilon$ prescription in the Lagrangian.

In the following, to discuss the imaginary part of this Lagrangian, we separately consider the two cases for the field ratio $B_I/E$ where $E\neq0$: (i) a rational and (ii) an irrational ratio.
These two cases require different treatments, particularly because the rational ratio can lead to coincident $E$ and $B_I$ poles, giving rise to the double poles.

When $B_I/E$ is irrational (i.e., $B_I/E\notin\mathbb{Q}$), the imaginary part of Eq.~\eqref{eq:L_scalar_EBI} is
\begin{widetext}
\begin{align}
\mathrm{Im}\,\mathcal{L}_{\text{scalar}}
=&~
\underbrace{\frac{e^2EB_I}{16\pi^2}
\sum_{n=1}^{\infty}
\frac{(-1)^{n-1}}{n}
\frac{
\exp\left(-\frac{n\pi m^2}{|eE|}\right)
}{
\sin\left(n\pi\frac{B_I}{E}\right)
}
}_\text{$E$-pole term}
+
\underbrace{
\frac{e^2EB_I}{16\pi^2}
\sum_{k=1}^{\infty}
\frac{(-1)^{k-1}}{k}
\frac{
\exp\left(-\frac{k\pi m^2}{|eB_I|}\right)
}{
\sin\left(k\pi\frac{E}{B_I}\right)
}
}_\text{$B_I$-pole term}
.
\label{eq:Im_L_scalar}
\end{align}
\end{widetext}
The first and second terms arise from the electric-field poles and the imaginary-magnetic-field poles, respectively.
Since now we assume an irrational ratio $B_I/E$, the electric-field poles
$s_n=n\pi/|eE|$ and the imaginary-magnetic-field poles
$s_k=k\pi/|eB_I|$ never coincide, and then the expression of the imaginary part can
be decomposed into the two simple-pole contributions.
In this sense, Eq.~\eqref{eq:Im_L_scalar} may be called the \emph{irrational formula} or \emph{noncoincident-pole formula}.

When a field ratio is rational (i.e., $B_I/E\in\mathbb{Q}$),
the imaginary part of Eq.~\eqref{eq:L_scalar_EBI} then becomes
\begin{widetext}
\begin{align}
\mathrm{Im}\,\mathcal{L}_{\text{scalar}}^{\text{rational}}
=&~
\underbrace{
\frac{e^2EB_I}{16\pi^2}
\sum_{\substack{n=1\\ q\nmid n}}^\infty
\frac{(-1)^{n-1}}{n}
\frac{
\exp\left(-\frac{n\pi m^2}{|eE|}\right)
}{
\sin\left(n\pi\frac{B_I}{E}\right)
}
}_\text{non-coincident $E$-pole term}
+\underbrace{
\frac{e^2EB_I}{16\pi^2}
\sum_{\substack{k=1\\ p\nmid k}}^\infty
\frac{(-1)^{k-1}}{k}
\frac{
\exp\left(-\frac{k\pi m^2}{|eB_I|}\right)
}{
\sin\left(k\pi\frac{E}{B_I}\right)
}}_\text{non-coincident $B_I$-pole term}
\nonumber\\
&+
 \underbrace{\frac{1}{16\pi^2} \sum_{j=1}^{\infty}
(-1)^{(p+q)j}
\exp\left(-\frac{\pi qj m^2}{|eE|}\right)  \left[
\frac{m^2|eE|}{qj}
+
\frac{(eE)^2}{\pi q^2j^2}
\right]. \hspace{60pt}}_\text{coincident-pole term}
\label{eq:Im_L_scalar_rational}
\end{align}
\end{widetext}
The first and second terms are similar to Eq.~\eqref{eq:Im_L_scalar}, but the restrictions $q\nmid n$ and $p\nmid k$ remove the {\it common poles}
of electric and imaginary magnetic sectors from the original sums, respectively\footnote{Here, $q\nmid n$ means that $n$ is not divisible by $q$: $q \nmid n \iff q \neq 0 \ \text{and}\ \frac{n}{q} \notin \mathbb{Z}$}, and
$p\nmid k$ means that $k$ is not divisible by $p$.
For common poles, the original pole indices $n$ and $k$ can be written as $n=qj$ and $k=pj$, respectively, where $j\in\mathbb{N}=\{1,2,3,\ldots\}$ is the common-pole index.
$p$ and $q$ are positive integer indices characterizing the field ratio
\begin{align}
\frac{B_I}{E}=\frac{p}{q},
\qquad
p,q\in\mathbb{N},
\qquad
\gcd(p,q)=1,
\label{eq:rational_ratio}
\end{align}
where $\gcd$ denotes the greatest common divisor.\footnote{Thus, $p$ and $q$ 
are coprime and $p/q$ is irreducible; for example, $2/3$ satisfies this condition, whereas $4/6$ does not.}
The common-pole contributions are contained in the third term, which arises from the double poles and
contains the two terms proportional to $m^2|eE|$ and $(eE)^2$.
Equation~\eqref{eq:Im_L_scalar_rational} may be called the \emph{rational formula}, which is composed of the first and second terms (i.e., the \emph{noncoincident-pole terms}) and the last term (the \emph{coincident-pole term}).

\subsubsection{Dirac QED}

Similarly, in Dirac QED,
\begin{align}
&\mathcal{L}_{\text{Dirac}}(E,B_I)
=-\frac{1}{8\pi^2}\int_0^\infty
\frac{ds}{s^3}e^{-m^2s} \nonumber\\
&~\times\biggl[
\frac{e^2EB_I s^2}
{\tan(eEs-i\epsilon)\tan(eB_Is-i\epsilon)}-1 + \frac{e^2(E^2+B_I^2)s^2}{3}
\biggr].
\label{eq:L_Dirac_EBI}
\end{align}
Thus, the effective Lagrangian is characterized by the product of two cotangent factors.
For $B_I/E\notin\mathbb{Q}$, its imaginary part is
\begin{align}
\mathrm{Im}\,\mathcal{L}_{\text{Dirac}}
=&~
\frac{e^2EB_I}{8\pi^2}
\sum_{n=1}^{\infty}\frac{1}{n}
\frac{\exp\left(-\frac{n\pi m^2}{|eE|}\right)}{\tan\left(n\pi\frac{B_I}{E}\right)}
\nonumber\\
&+
\frac{e^2EB_I}{8\pi^2}
\sum_{k=1}^{\infty} \frac{1}{k}
\frac{
\exp\left(-\frac{k\pi m^2}{|eB_I|}\right)}{\tan\left(k\pi\frac{E}{B_I}\right)}.
\label{eq:Im_L_Dirac}
\end{align}
For $B_I/E=p/q\in\mathbb{Q}$, we obtain
\begin{align}
&\mathrm{Im}\,\mathcal{L}_{\text{Dirac}}^{\text{rational}}\nonumber\\
&=
\frac{e^2EB_I}{8\pi^2}
\sum_{\substack{n=1\\q\nmid n}}^\infty \frac{1}{n}
\frac{\exp\left(-\frac{n\pi m^2}{|eE|}\right)}{\tan\left(n\pi\frac{B_I}{E}\right)}
\nonumber\\
&~+
\frac{e^2EB_I}{8\pi^2}
\sum_{\substack{k=1\\p\nmid k}}^\infty \frac{1}{k}
\frac{\exp\left(-\frac{k\pi m^2}{|eB_I|}\right)}{\tan\left(k\pi\frac{E}{B_I}\right)}
\nonumber\\
&~- \frac{1}{8\pi^2}
\sum_{j=1}^{\infty}
\exp\left(-\frac{\pi qj m^2}{|eE|}\right)
\left[
\frac{m^2|eE|}{qj}
+
\frac{(eE)^2}{\pi q^2j^2}
\right].
\label{eq:Im_L_Dirac_rational}
\end{align}

\subsubsection{Special cases}

Here, we discuss two useful cases.
First, in the limit of $E\to0$, i.e., in pure imaginary magnetic fields, only the the second terms in each formula survives:
\begin{align}
\mathrm{Im}\,\mathcal{L}_{\mathrm{scalar}}(0,B_I)
&=\frac{(eB_I)^2}{16\pi^3}
\sum_{n=1}^{\infty}\frac{(-1)^{n-1}}{n^2}
e^{-\frac{n\pi m^2}{|eB_I|}},
\label{eq:pure_BI_scalar}\\
\mathrm{Im}\,\mathcal{L}_{\mathrm{Dirac}}(0,B_I)
&=\frac{(eB_I)^2}{8\pi^3}
\sum_{n=1}^{\infty}\frac{1}{n^2}
e^{-\frac{n\pi m^2}{|eB_I|}}.
\label{eq:pure_BI_dirac}
\end{align}
These expressions are the same functional form as the ordinary pure electric results after $|eE|\leftrightarrow|eB_I|$.

Second, when the field strengths are equal, $|eE|=|eB_I|$, we have $p=q=1$.
Then, in Eqs.~\eqref{eq:Im_L_scalar_rational} and \eqref{eq:Im_L_Dirac_rational}, every pole is a common pole, and the two noncoincident-pole terms vanish.
Then, only the third term survives:
\begin{align}
\mathrm{Im}\,\mathcal{L}_{\mathrm{scalar}}
&=\frac{|eB_I|}{16\pi^2}
\left[m^2\,\operatorname{Li}_1(z)
+\frac{|eB_I|}{\pi}\operatorname{Li}_2(z)\right],
\label{eq:diagonal_scalar}\\
\mathrm{Im}\,\mathcal{L}_{\mathrm{Dirac}}
&=-\frac{|eB_I|}{8\pi^2}
\left[m^2\,\operatorname{Li}_1(z)
+\frac{|eB_I|}{\pi}\operatorname{Li}_2(z)\right],
\label{eq:diagonal_dirac}
\end{align}
where we defined $z=\exp(-\pi m^2/|eB_I|)$.
$\operatorname{Li}_1(z)=\sum_{j=1}^\infty \frac{z^j}{j}$ and $\operatorname{Li}_2(z)=\sum_{j=1}^\infty \frac{z^j}{j^2}$, where $|z|\leq 1$, are the polylogarithm functions.
These formulas are the same as that in Ref.~\cite{Dunne:2002qg}, obtained from the effective Lagrangian~\cite{Dunne:2001pp,Dunne:2002qf,Dunne:2002qg,Dunne:2002ta,Dunne:2003tr} for a Euclidean-self-dual background.
In this sense, $|eE|=|eB_I|$ may be called the \emph{Euclidean-self-dual point}. 
Also, there is an exact relation between the two formulas for the scalar and Dirac cases:\footnote{This relation reflects a supersymmetric quantum-mechanical structure between the Klein-Gordon and Dirac operators in the self-dual background~\cite{DAdda:1977sqj}.}
\begin{align}
\mathrm{Im}\,\mathcal{L}_{\mathrm{Dirac}}
=-2\,\mathrm{Im}\,\mathcal{L}_{\mathrm{scalar}}
\qquad (|eE|=|eB_I|).
\label{eq:diagonal_relation}
\end{align}

\subsection{Complex Landau levels}

In general, a physical interpretation in terms of Landau levels is useful for understanding the physics in real magnetic fields.
For a real magnetic field, the transverse motion of charged particles is quantized into discrete Landau levels, and the vacuum decay rate can be expressed as a sum over
these levels (see Appendix~\ref{app:landau}).
For an imaginary magnetic field, conventional Landau levels are no longer applicable, but they can be extended to a complex Landau-level spectrum.
We here examine the Landau-level structure of the Schwinger effect in the presence of an imaginary magnetic field.
 
For scalar QED, the corresponding effective squared masses are real
\begin{align}
M_\ell^2(B)
=
m^2+(2\ell+1)|eB|,
\qquad \ell=0,1,2,\ldots .
\end{align}
Using the analytic continuation $B\to iB_I$,\footnote{Notice that this analytic continuation should not be understood as $|eB|\to|ieB_I|$.
Instead, $|eB|=eB \to ieB_I$ for $eB>0$ while $|eB|=-eB \to -ieB_I=i|eB_I|$ for $eB<0$.} they become
complex,
\begin{align}
M_\ell^2(B_I)
=
m^2+i(2\ell+1)|eB_I|.
\label{eq:complex_LL_scalar}
\end{align}
We refer to this spectrum as \emph{complex Landau levels}, because the eigenvalues are complex.
The imaginary parts of the eigenvalues reflect the resonant nature of the transverse motion, which can be understood in terms of the Gamow states of an inverted harmonic oscillator~\cite{Barton:1984ey,Subramanyan:2020fmx}.

Using the complex Landau levels, the electric-pole contribution, i.e., the first term of Eq.~\eqref{eq:Im_L_scalar}, can be rewritten as an Abel-regularized Landau-level representation (for a detailed derivation, see Appendix~\ref{app:landau_imag}),
\begin{align}
\mathrm{Im}\mathcal{L}_{\mathrm{scalar}}^{(E)}
&=
\lim_{\eta\to0^+}
\mathrm{Re}\Bigg\{
\frac{i e^2EB_I}{8\pi^2}
\sum_{\ell=0}^{\infty}
e^{-(2\ell+1)\eta}
\nonumber\\
&\qquad\times
\ln\left[
1+
\exp\left(
-\frac{\pi M_\ell^2(B_I)}{|eE|}
\right)
\right]
\Bigg\}.
\label{eq:complex_LL_scalar_rate}
\end{align}
Taking the real part is required to obtain the imaginary part of complex quantity.
The overall factor $i$ is regarded as the analytic continuation of the Landau degeneracy factor $|eB|/2\pi$.
$\eta>0$ is the Abel regulator.
In an imaginary magnetic field, the exponential factors for higher Landau levels no longer provide exponential suppression but instead become oscillatory.
Then, the Abel regulator is needed to ensure convergence of the Landau-level sum.
By summing all the Landau levels and taking the limit $\eta\to0$, Eq.~\eqref{eq:complex_LL_scalar_rate} agrees with the first term of Eq.~\eqref{eq:Im_L_scalar}.

For Dirac QED, the analytically continued effective squared
masses are
\begin{align}
M_\ell^2(B_I)
=
m^2+2i\ell |eB_I|,
\qquad \ell=0,1,2,\ldots .
\label{eq:complex_LL_dirac}
\end{align}
The difference from the spectrum in scalar QED originates from the spin structure of Dirac field, and, in the nonrelativistic picture, its imaginary shift can be interpreted as the imaginary Zeeman effect.

The electric-pole contribution, i.e., the first term of Eq.~\eqref{eq:Im_L_Dirac}, can be written as
\begin{align}
\mathrm{Im}\mathcal{L}_{\mathrm{Dirac}}^{(E)} &=
\lim_{\eta\to0^+}
\mathrm{Re}\Bigg\{
-\frac{i e^2EB_I}{8\pi^2}
\sum_{\ell=0}^{\infty}
g_\ell e^{-2\ell\eta}
\nonumber\\
&\qquad\times
\ln\left[
1-
\exp\left(
-\frac{\pi M_\ell^2(B_I)}{|eE|}
\right)
\right]
\Bigg\}.
\label{eq:complex_LL_dirac_rate}
\end{align}
The overall minus sign and the $1-\exp$ form in the logarithmic function reflect Fermi statistics.
The degeneracy factor $g_l$ comes from the spin structure of Landau levels in Dirac QED.
The lowest Landau level $\ell=0$ has no spin degeneracy, $g_0=1$, and higher Landau levels $\ell\geq1$ have the spin degeneracy, $g_{\ell}=2$.

Note that our current purpose is the comparison of the Landau level structures in real and imaginary magnetic fields.
Therefore, in this section, we focused on the Landau-level decompositions of electric poles.
Similarly, for the imaginary magnetic poles, we can formally discuss the discrete spectrum depending on the electric field.

In the next section, we numerically investigate the behaviors of individual complex Landau-level contributions and their sum.

\section{Numerical results and discussion}
\label{sec:results}

In this section, we present numerical results for the imaginary part of the effective Lagrangian in the presence of an imaginary magnetic field.
We focus on four characteristic features: (i) its dependence on $B_I$, (ii) the separate contributions from the electric and imaginary-magnetic poles, (iii) the convergence of the corresponding infinite sums, and (iv) the contributions of the complex Landau levels.

\begin{figure}[b]
\centering
\includegraphics[width=0.9\columnwidth]{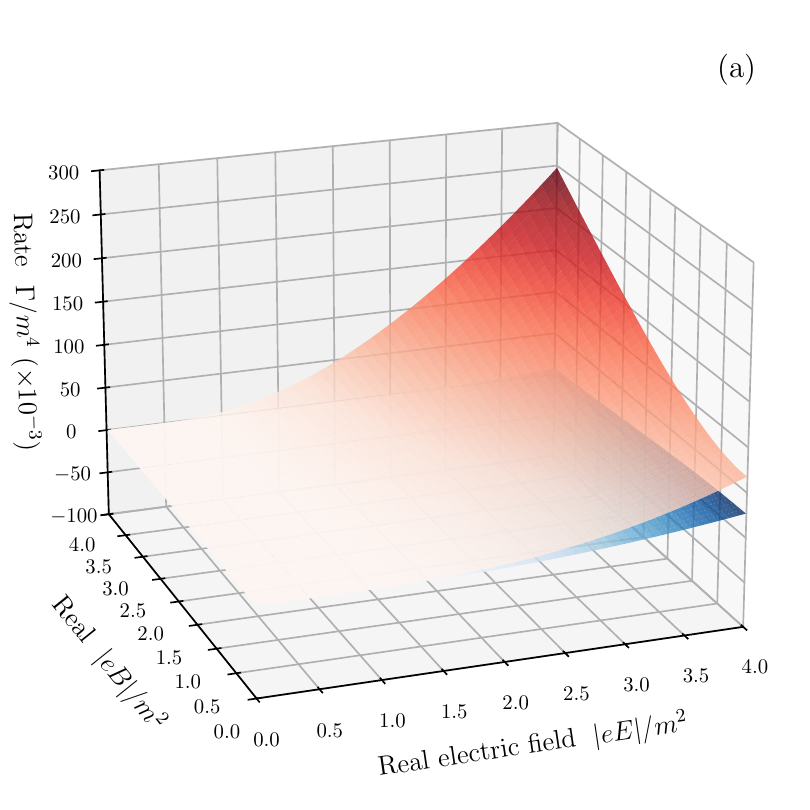}
\includegraphics[width=0.9\columnwidth]{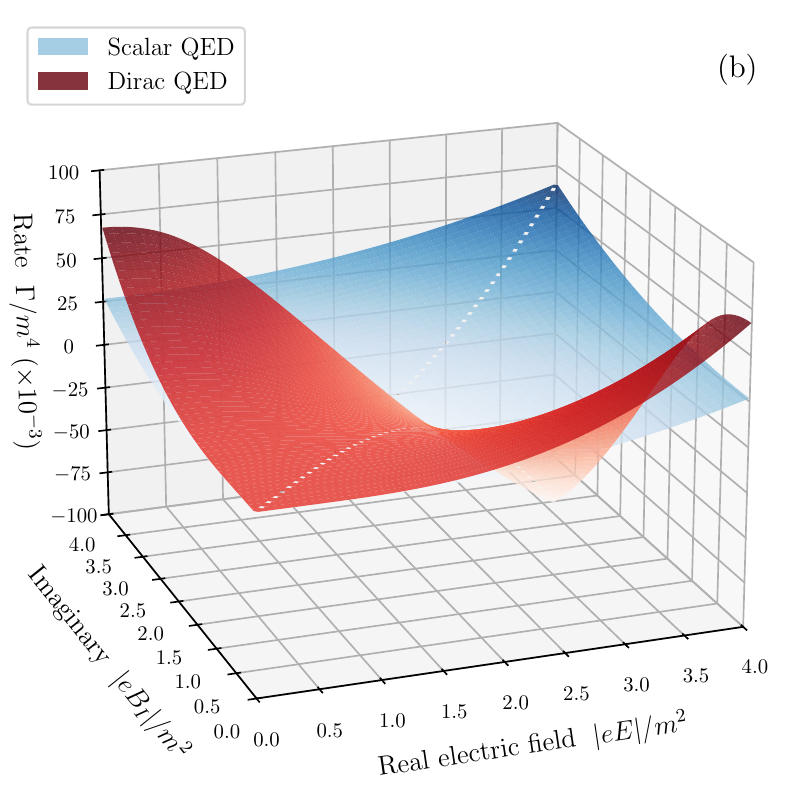}
\caption{The imaginary part of the effective
Lagrangian, $\Gamma=2\mathrm{Im}\mathcal{L}$, for scalar and
Dirac QED as functions of the electric and magnetic fields:
(a) real magnetic field $B$ and (b) imaginary magnetic field
$B_I$.}
\label{fig:imag-EB-3D}
\end{figure}

\subsection{Imaginary part of the effective Lagrangian}

In Fig.~\ref{fig:imag-EB-3D}(a) and (b), we plot the imaginary part of the one-loop
effective Lagrangian on both $(E,B)$ and $(E,B_I)$ planes, respectively, where the vertical axis shows the dimensionless quantity $\Gamma/m^4$, with $\Gamma \equiv 2\mathrm{Im}\mathcal{L}$.
In real magnetic fields shown in Fig.~\ref{fig:imag-EB-3D}(a), $\Gamma$ vanishes along the $E=0$
boundary for any value of $B$.
It is well known that a static real magnetic field does not generate an imaginary part of the effective Lagrangian by itself but modifies the contribution induced by a nonzero electric field.
At a fixed nonzero $E$, as $B$ increases, $\Gamma$ is suppressed in scalar QED and enhanced in Dirac QED.
This property reflects the different Landau-level structures of the scalar and Dirac fields (see Appendix~\ref{app:landau} for details).

In imaginary magnetic fields, the behavior changes qualitatively, as shown in Fig.~\ref{fig:imag-EB-3D}(b).
We find that the dependence on $E$ and $B_I$ becomes symmetric under their interchange, and a nonzero $B_I$ generates a finite imaginary part even along the $E=0$ boundary.
The ordinary pure-electric result is recovered at $B_I=0$.
This {\it symmetric behavior} is clear from the formulas in Eqs.~\eqref{eq:Im_L_scalar}, \eqref{eq:Im_L_scalar_rational}, \eqref{eq:Im_L_Dirac}, and \eqref{eq:Im_L_Dirac_rational}.

\begin{figure}[b]
\centering
\includegraphics[width=1\columnwidth]{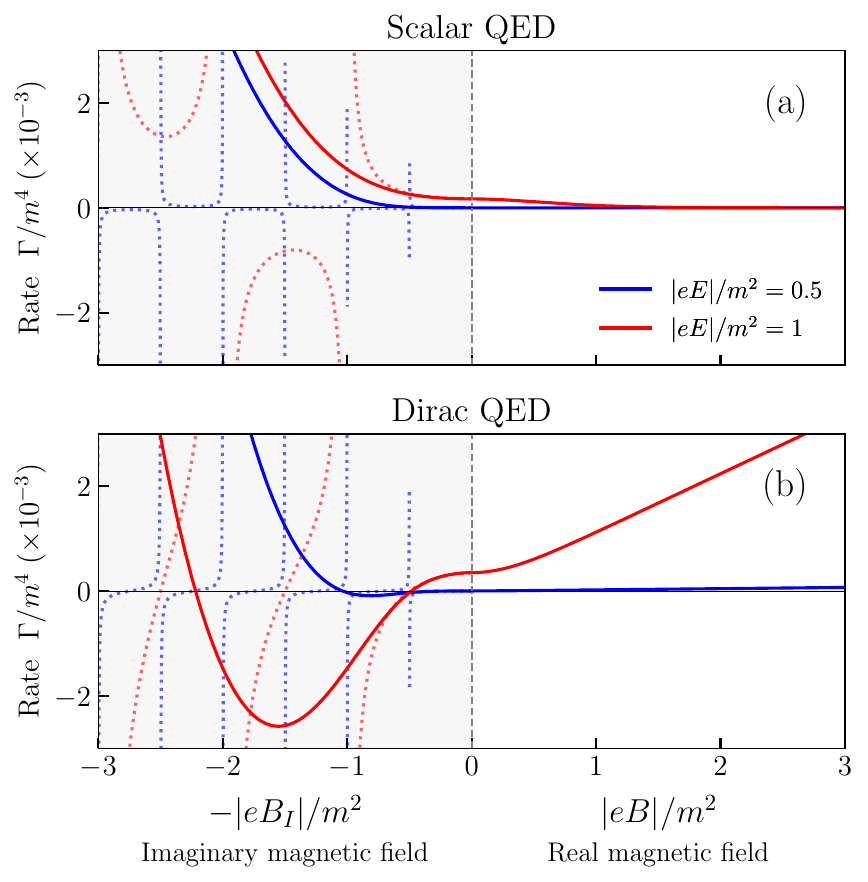}
\caption{The imaginary part of the effective
Lagrangian for (a) scalar and (b) Dirac QED as a function of the
imaginary (left) and real (right) magnetic fields at fixed
$|eE|/m^2=0.5$ (blue) and $1$ (red). In the imaginary-field region, the solid lines show the full result,
including both $E$- and $B_I$-pole contributions, while the
dotted lines show the $E$-pole contribution alone.
A real magnetic field suppresses the scalar result and enhances the Dirac result, whereas the imaginary-field region exhibits nonmonotonic and, for Dirac QED, sign-changing behavior.}
\label{fig:imag-EB-cont}
\end{figure}

To examine the field dependence more closely,
Fig.~\ref{fig:imag-EB-cont} shows $\Gamma$ at fixed electric-field
strengths, $|eE|/m^2=0.5$ and $1$. The real and imaginary
magnetic-field regions are displayed on the positive and negative
sides of the horizontal axis, respectively. In the real-field region,
the scalar result decreases monotonically as $B$ increases, whereas
the Dirac result increases. In the imaginary-field region, the
dependence on $B_I$ is considerably more intricate: the scalar result increases rapidly, while the Dirac result is nonmonotonic:
it is initially negative but becomes positive as $B_I$ increases.
These differences become more pronounced at larger electric-field strengths.
We note that, near $B_I=0$, the full results (solid lines) approach
the $E$-pole contributions (dotted lines), consistent
with the fact that only $E$ poles contribute for real
magnetic fields. As $B_I$ increases, the additional
$B_I$-pole contributions become significant,
leading to increasing deviations.

The qualitative difference between real and imaginary magnetic fields can be understood  from the proper-time
structure.
For a real magnetic field, the $B$-dependence is
described by hyperbolic functions that introduce no additional poles
on the positive real proper-time axis. For an imaginary magnetic field, these
functions become trigonometric and generate an additional family of
poles, together with the conventional electric poles.
The interplay between these two pole families produces the
symmetric dependence on $E$ and $B_I$ in
Fig.~\ref{fig:imag-EB-3D}(b), as well as the nonmonotonic and
sign-changing behavior shown in Fig.~\ref{fig:imag-EB-cont}.

\begin{figure}[b]
\centering
\includegraphics[width=1\columnwidth]{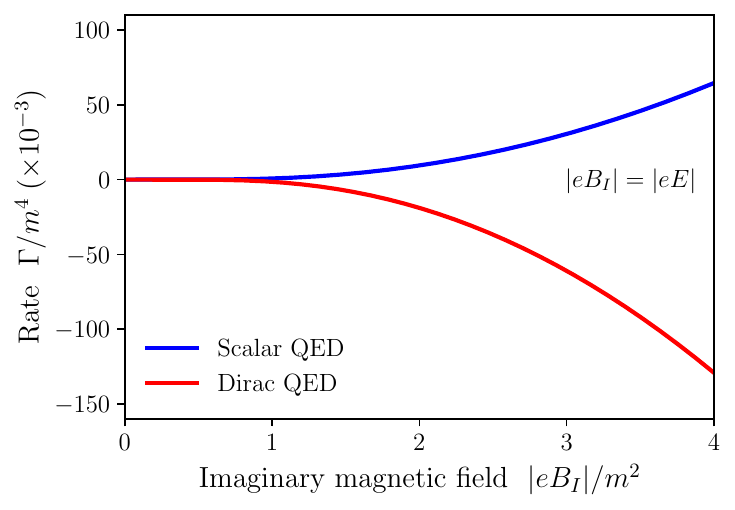}
\caption{The imaginary part of the effective
Lagrangian for scalar and Dirac QED along the line of
$|eE|=|eB_I|$. The scalar result is positive and increases with
the field strength, whereas the Dirac result is negative and
satisfies $\Gamma_{\mathrm{Dirac}}=-2\Gamma_{\mathrm{scalar}}$.}
\label{fig:imag-EB-diag}
\end{figure}

Figure~\ref{fig:imag-EB-diag} shows the result where the electric and imaginary magnetic fields are varied simultaneously
along the line of $|eE|=|eB_I|$, which corresponds to Euclidean-self-dual field.
We find that the scalar result is positive and increases with the common field strength,
whereas the Dirac result is negative and decreases.
As analytically derived in Eq.~\eqref{eq:diagonal_relation}, the two results
satisfy the exact relation
$\Gamma_{\mathrm{Dirac}}=-2\Gamma_{\mathrm{scalar}}$.

The physical interpretation of the imaginary part $\Gamma$ of the effective Lagrangian in an imaginary magnetic field requires some care.
For the conventional Hermitian electromagnetic field, $\Gamma$ characterizes the vacuum-persistence probability~\cite{Schwinger:1951nm}, $\left|e^{iVT\mathcal{L}}\right|^2=e^{-2VT \mathrm{Im}\mathcal{L}}=e^{-VT\Gamma}$, where the probabilistic interpretation is based on the unitarity of the time evolution and cannot be directly applied to a non-Hermitian background.
In the non-Hermitian case, the sign of $\Gamma$ determines whether the vacuum amplitude decays or grows, and its magnitude determines the strength of this behavior.
Thus, $\Gamma$ should not be treated as the \emph{probabilistic vacuum decay rate} but can be regarded as \emph{a rate of vacuum amplitude decay} ($\Gamma>0$) or \emph{amplification} ($\Gamma<0$).\footnote{
When we define the squared vacuum amplitude $|\mathcal{A}_{\rm vac}|^2=e^{-VT\Gamma}$, $|\mathcal{A}_{\rm vac}|^2 =P_{\rm vac}$ in the Hermitian case, whereas $|\mathcal{A}_{\rm vac}|^2 \neq P_{\rm vac}$ in the non-Hermitian case.}

Although in this subsection we found that the total imaginary part appears smooth, each contribution from the electric- or imaginary-magnetic-pole terms exhibits oscillatory and
singular structures as functions of $B_I/E$ because of the trigonometric factors, such as $1/\sin(n\pi B_I/E)$ and $1/\sin(k\pi E/B_I)$ in scalar QED and the cotangent factors in Dirac QED.
Therefore, we can expect that a cancellation mechanism works.
In the next subsection, we examine the two-pole contributions individually and investigate how their singularities are canceled.

\begin{figure*}[t]
\centering
\includegraphics[width=2.0\columnwidth]{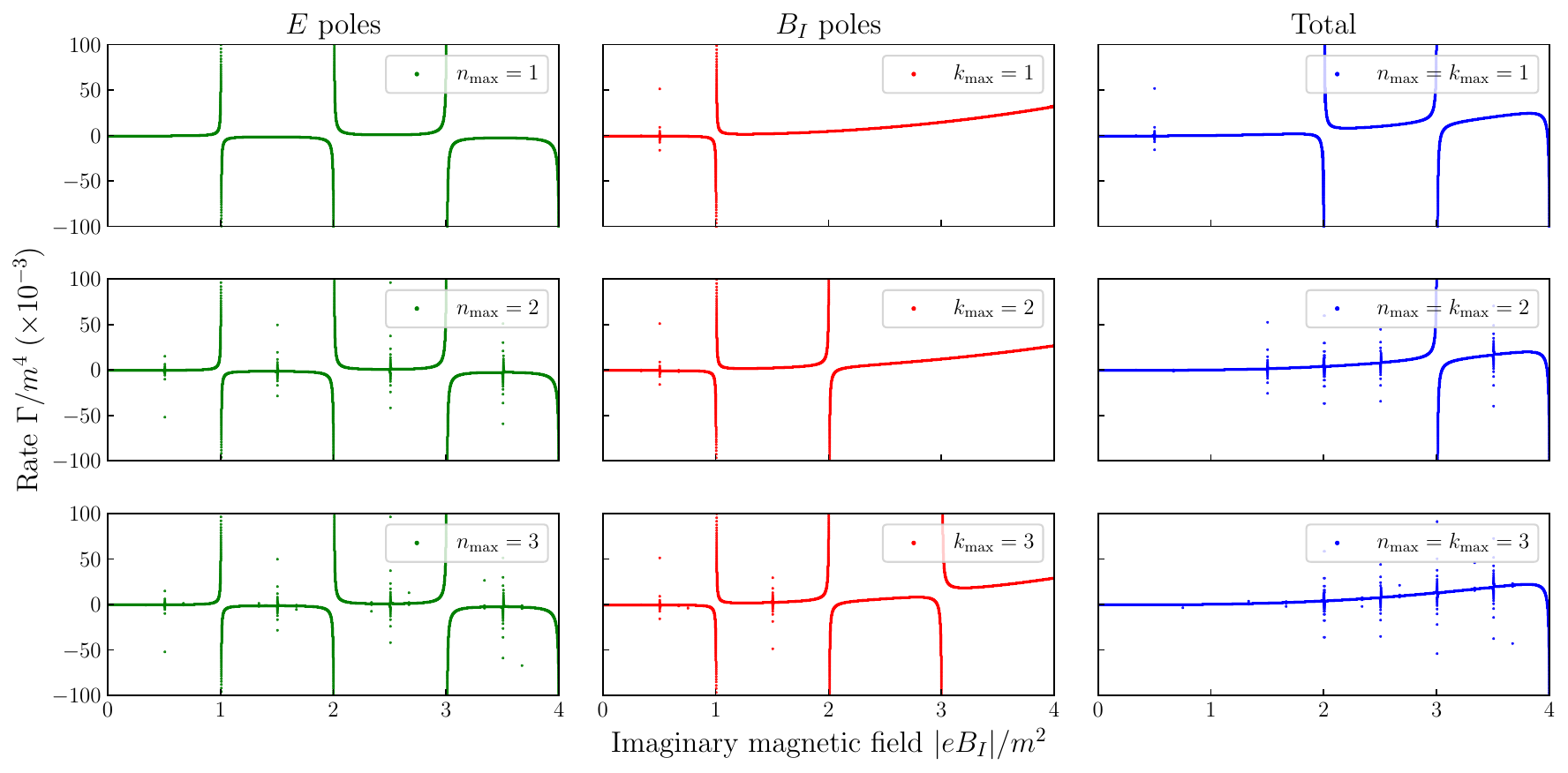} 
    \caption{Electric-pole (left), imaginary-magnetic-pole (middle), and
    total (right) contributions to the imaginary part of the effective
    Lagrangian in scalar QED as functions of the imaginary magnetic
    field strength $|eB_I|/m^2$ at fixed $|eE|/m^2=1$.
    The upper, middle, and lower rows correspond to truncation orders
    $n_{\mathrm{max}}(k_{\mathrm{max}})=1$, $2$, and $3$, respectively.}
    \label{fig:imag-EB-scalar-poles}
\end{figure*}

\begin{figure*}[t]
\centering
\includegraphics[width=2.0\columnwidth]{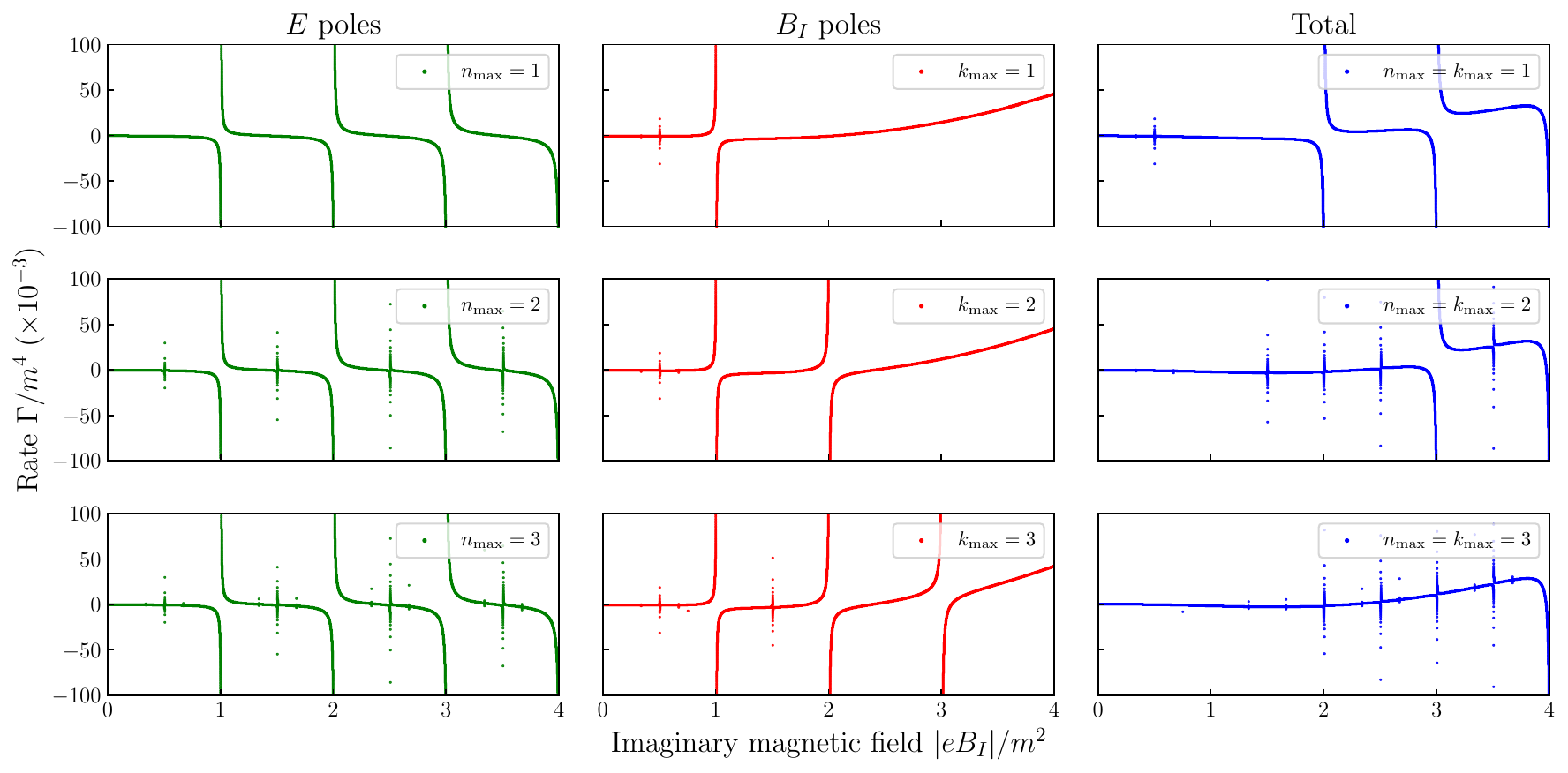}
\caption{Same as Fig.~\ref{fig:imag-EB-scalar-poles}, but for Dirac
QED.}
\label{fig:imag-EB-dirac-poles}
\end{figure*}

\subsection{Electric- and imaginary-magnetic-pole contributions}

Here, we examine the
two-pole contributions separately.
For scalar QED, the imaginary part can be written schematically as
\begin{align}
\mathrm{Im}\,\mathcal{L}_{\mathrm{scalar}}
=
\mathrm{Im}\,\mathcal{L}^{(E)}
+
\mathrm{Im}\,\mathcal{L}^{(B_I)},
\end{align}
where the first and second terms arise from the electric and
imaginary-magnetic poles, respectively.

The scalar-QED results are shown in
Fig.~\ref{fig:imag-EB-scalar-poles} for truncation orders
$n_{\mathrm{max}}=k_{\mathrm{max}}=1$, $2$, and $3$.
The electric-pole contribution contains the factor $1/\sin(n\pi B_I/E)$, so that singularities are located at $nB_I/E\in\mathbb{Z}$.
For $n=1$, the poles occur at integer values of $B_I/E$,
while for $n\ge2$, additional poles appear at fractional values of $B_I/E$.
In contrast, the imaginary-magnetic-pole contribution contains $1/\sin(k\pi E/B_I)$, and singularities are located at $kE/B_I\in\mathbb{Z}$.
In the left and middle columns of Fig.~\ref{fig:imag-EB-scalar-poles}, we separately plot the contribution of each pole family as a function of $B_I/E$, where we find that $\Gamma$ has singularities and oscillates.
With increasing the truncation order, additional singularities appear.

As shown in the right column of Fig.~\ref{fig:imag-EB-scalar-poles}, the total contribution exhibits a different behavior
due to the interplay between the two pole families.
For $n_{\mathrm{max}}=k_{\mathrm{max}}=1$, the singularity near $B_I/E=1$ is canceled between the contributions from the electric and imaginary magnetic poles, while other singularities near $B_I/E=1/2$ from the magnetic poles and $B_I/E=2,3,\dots$ from the electric poles remain.
Similarly, for $n_{\mathrm{max}}=k_{\mathrm{max}}=2$, the singularities near $B_I/E=1/2$ and $2$ are canceled.
For $n_{\mathrm{max}}=k_{\mathrm{max}}=3$, the singularities near $B_I/E=1/3$, $2/3$, $3/2$, and $3$ are canceled.
Due to this cancellation mechanism, when $n_{\mathrm{max}}$ is sufficiently large, the singular behavior in $\Gamma$ is suppressed, and $\Gamma$ becomes a smooth function.

The corresponding Dirac-QED results are displayed in
Fig.~\ref{fig:imag-EB-dirac-poles}.
The two pole families occur at the same field ratios
as in scalar QED, but their residues differ because
the Dirac effective Lagrangian contains cotangent factors, instead of the cosecant factors of the scalar case.
Its cancellation mechanism is similar to the scalar case. 

\subsection{Convergence of the pole sums}

\begin{figure}[b]
    \centering
    \includegraphics[width=\columnwidth]{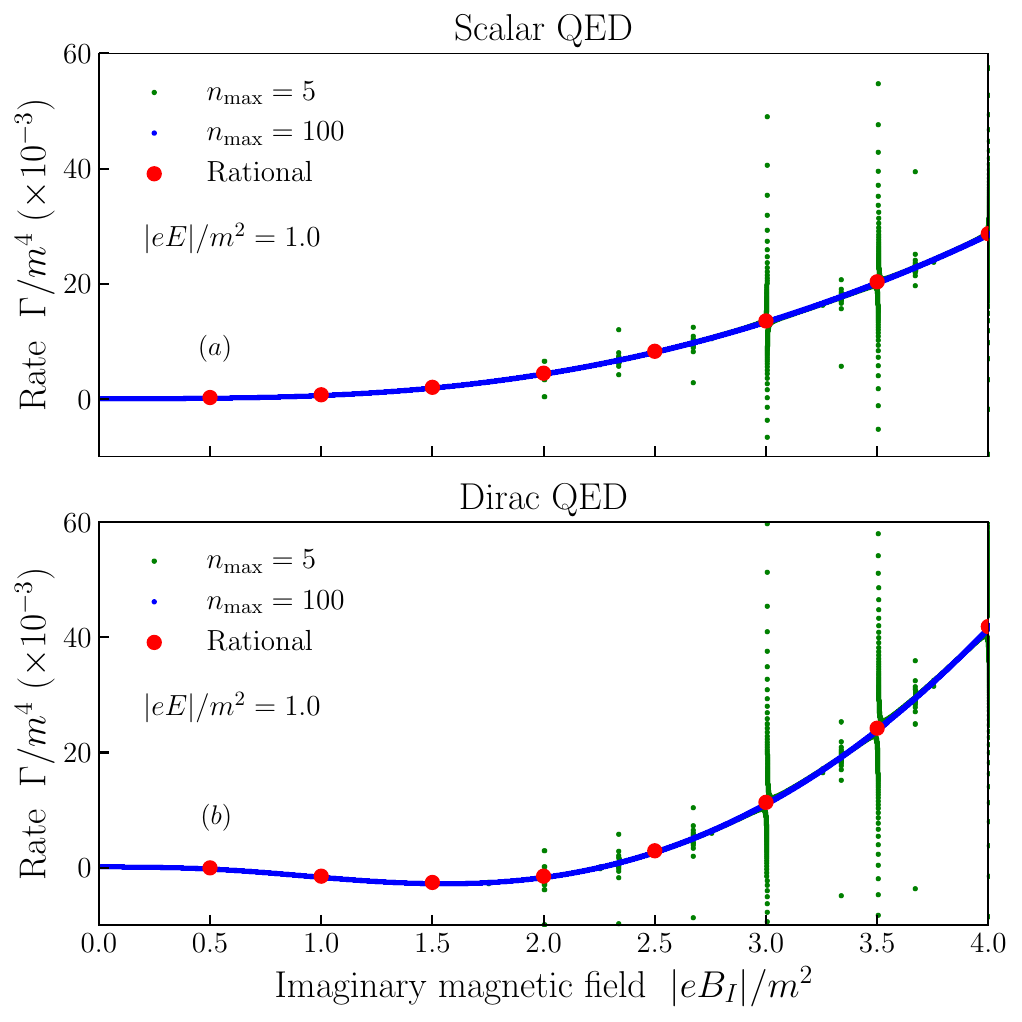}
    \caption{Convergence of $\Gamma=2\,\mathrm{Im}\,\mathcal{L}$
    with increasing pole-sum truncation $n_{\mathrm{max}}$
    for (a) scalar and (b) Dirac QED at $|eE|/m^2=1$.
    Green and blue points correspond to $n_{\mathrm{max}}=k_{\mathrm{max}}=5$
    and $100$, respectively, obtained from the irrational formulas, Eqs.~\eqref{eq:Im_L_scalar} and \eqref{eq:Im_L_Dirac}. Red circles denote results at
    selected rational field ratios in the rational formulas \eqref{eq:Im_L_scalar_rational} and \eqref{eq:Im_L_Dirac_rational}, where $n_{\mathrm{max}}$ is sufficiently large.}
    \label{fig:imag-EB-convergence}
\end{figure}

Figure~\ref{fig:imag-EB-convergence} demonstrates the convergence of $\Gamma$ with increasing the truncation order, where we compare the results truncated at $n_\mathrm{max}=k_\mathrm{max}=$ 5 and $100$ in the noncoincident-pole formulas in Eqs.~\eqref{eq:Im_L_scalar} and \eqref{eq:Im_L_Dirac}.
At low truncation order $n_\mathrm{max}=5$, the resulting rates exhibit pronounced spike-like structures near the poles of the individual terms. These structures reflect the incomplete cancellation between contributions from the two pole families.

Increasing the truncation order  modifies the behavior near coincident poles.
In particular, the large spikes in low-order truncations are reduced near certain rational field ratios as the corresponding
electric- and imaginary-magnetic-pole contributions are combined.
For nonsingular field ratios, the partial sums approach the
large-$n$ result.

This convergence behavior is qualitatively similar for scalar and
Dirac QED, although their pole contributions differ through the their overall coefficients, sign structure, and trigonometric factors in the respective
expressions.
The results emphasize that the electric- and
imaginary-magnetic-pole sectors must be considered together. 

At exactly rational field ratios, $B_I/E=p/q$, the noncoincident-pole formulas in Eqs.~\eqref{eq:Im_L_scalar} and \eqref{eq:Im_L_Dirac} are not applicable because the electric and imaginary-magnetic poles overlap.
Then, the common poles must be excluded from the two simple-pole sums, and their contributions are evaluated from the corresponding double-pole residues, as the third term in the formulas in Eqs.~\eqref{eq:Im_L_scalar_rational} and \eqref{eq:Im_L_Dirac_rational}.
The red circles in Fig.~\ref{fig:imag-EB-convergence} show the results obtained using these formulas.
They agree with the limiting values obtained from nearby irrational values of $B_I/E$,
which confirms the consistency between the rational and irrational formulas.

\subsection{Complex Landau-level structure}

\begin{figure}[b]
    \centering
    \includegraphics[width=\columnwidth]
    {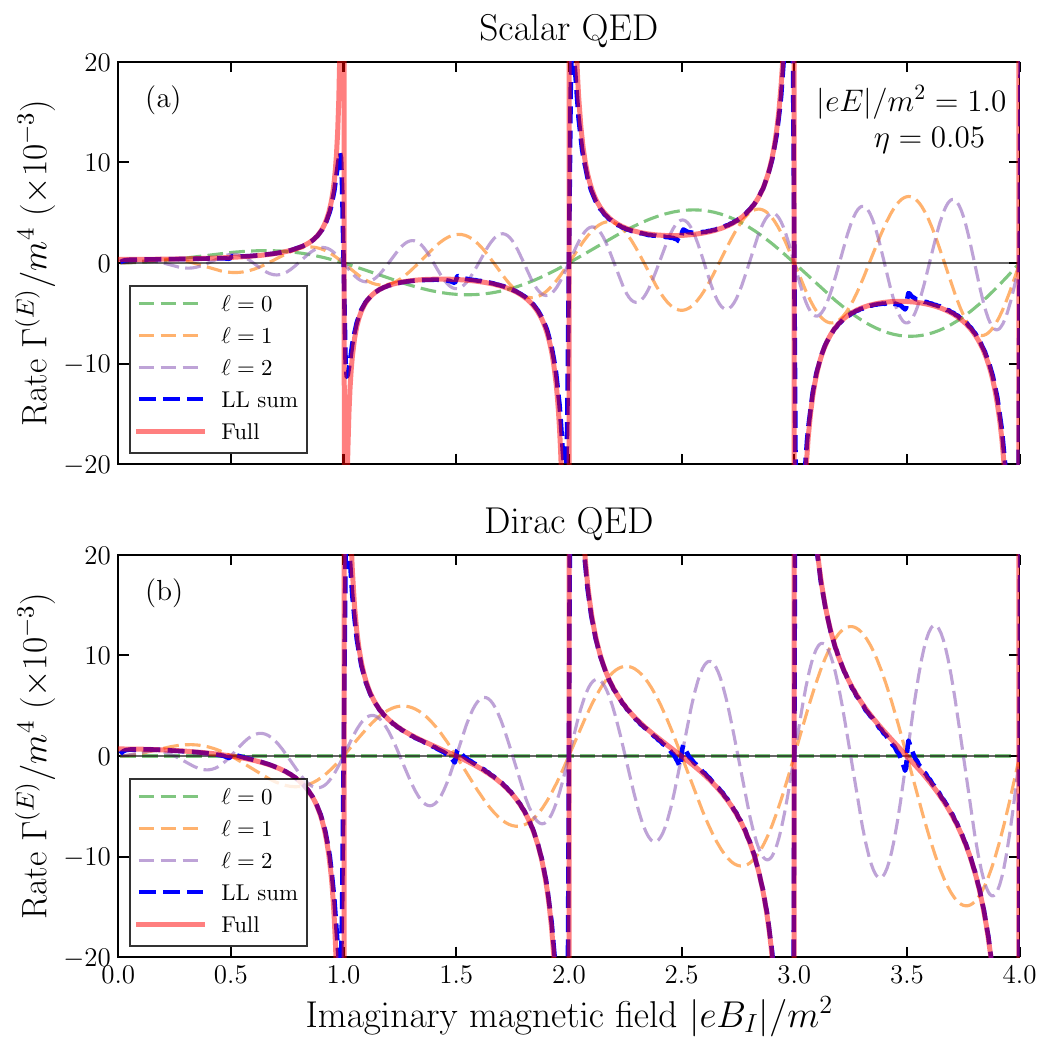}
    \caption{Landau level decomposition of electric-pole contribution, $\Gamma^{(E)}\equiv2\,\mathrm{Im}\,
    \mathcal{L}- \Gamma^{(B_I)}$, as a function of the imaginary
    magnetic field strength $|eB_I|/m^2$ for
    (a) scalar and (b) Dirac QED at fixed
    $|eE|/m^2=1$. The Landau-level sum is truncated at $\ell_{\max}=50$, and the full result with $n_{\mathrm{max}}=1$ is shown for comparison.
    Only the electric-pole contribution is included whereas the imaginary-magnetic-pole contribution is not included.}
    \label{fig:imag-EB-complex-landau}
\end{figure}

Figure~\ref{fig:imag-EB-complex-landau} shows the
contributions of the lowest three complex Landau levels
and their regulated sum, obtained from Eqs.~\eqref{eq:complex_LL_scalar_rate} and \eqref{eq:complex_LL_dirac_rate} for scalar and Dirac QED.
Here, we fix the Abel regulator at $\eta=0.05$ for illustration. As $\ell_{\max}$ increases, the results become closer to the full result. A smaller $\eta$ requires a larger $\ell_{\max}$ to obtain similar results.
Each individual level exhibits oscillatory behavior as $|eB_I|$ increases, reflecting the phases generated by their complex effective squared masses. Higher Landau levels oscillate more rapidly because their phases are proportional to
$(2\ell+1)B_I/E$ in scalar QED and $2\ell B_I/E$
in Dirac QED.

In the two
theories, the lowest Landau level ($\ell=0$) behaves differently.
In scalar QED, the effective squared mass contains an imaginary shift and produces an oscillatory contribution. In Dirac QED, the lowest
Landau-level is a real spectrum, $M_{0,\mathrm{Dirac}}^2=m^2$.
Then, its contribution to $\Gamma^{(E)}$ vanishes, as shown by the green dashed curve in Fig.~\ref{fig:imag-EB-complex-landau}(b).
The mathematical reason is understood from Eqs.~\eqref{eq:complex_LL_scalar_rate} and \eqref{eq:complex_LL_dirac_rate}:
Since $\Gamma^{(E)}$ is obtained by taking the real part of a complex integrand with the overall imaginary unit $i$, a real Landau level does not contribute to  $\Gamma^{(E)}$.
Thus, the nonzero $\Gamma^{(E)}$ is induced by imaginary shifts in the complex Landau-level spectrum.

The blue dashed lines in Fig.~\ref{fig:imag-EB-complex-landau} represent the results obtained by summing over Landau levels up to $\ell=50$.
Although the contribution from each Landau level is a sinusoidal-like oscillation, the sum deforms the functional forms.
As the number of Landau levels increases, the sum gradually approaches the full results of $\Gamma^{(E)}$, i.e., the sharp divergences seen in the left column of Figs.~\ref{fig:imag-EB-scalar-poles} and \ref{fig:imag-EB-dirac-poles}.
 
Thus, a decomposition picture in terms of complex Landau levels is formally possible.
However, unlike the case of a real magnetic field, the higher Landau levels are not exponentially suppressed. Consequently, an LLL-dominance picture cannot be straightforwardly applied. Nevertheless, the complex Landau-level decomposition may provide a useful qualitative picture for interpreting the full electric-pole result as a sum of individual sinusoidal-like modes.

\section{Conclusion and outlook}
\label{sec:conclusion}

In this work, we have investigated the Schwinger effect in
the presence of an imaginary magnetic field parallel to a
real electric field. We construct the one-loop effective
Lagrangians of scalar and Dirac QED, Eqs.~\eqref{eq:L_scalar_EBI} and \eqref{eq:L_Dirac_EBI}, respectively, which are obtained from the analytic continuation $B\to iB_I$.
This continuation
transforms the hyperbolic factors as a function of real magnetic field into
trigonometric functions, which leads to additional poles on the real proper-time axis.

Our main findings are the imaginary parts of the effective Lagrangians, Eqs.~\eqref{eq:Im_L_scalar}, \eqref{eq:Im_L_scalar_rational}, \eqref{eq:Im_L_Dirac}, and \eqref{eq:Im_L_Dirac_rational}.
These formulas contain contributions from both the conventional electric-field
poles, $s_n=n\pi/|eE|$, and the imaginary-magnetic-field
poles, $s_k=k\pi/|eB_I|$. Their interplay gives rise to
nontrivial field dependence and cancellations between the
two contributions. When $B_I/E$ is rational, some poles from the two families coincide and form double poles.
At these field ratios, the separate simple-pole expressions are no longer applicable, and the common poles must be treated carefully.

Natural extensions of the present study are as follows:
\begin{itemize}
\item It would be of interest to study a complex magnetic field, $B=B_R+iB_I$, with both real and imaginary components.
Such an extension would allow us to investigate how the proper-time pole structure and complex Landau levels evolve as the magnetic field is continuously varied in the complex plane.
\item It would be interesting to reformulate or rederive our results using other theoretical approaches, including the operator formalism~~\cite{Nikishov:1969tt}, semiclassical approaches such as the worldline instanton method~\cite{Affleck:1981bma}, holographic approaches~\cite{Semenoff:2011ng}, and resurgence methods~\cite{Florio:2019hzn}.
\item It would also be interesting to extend the study to the QCD Schwinger effect, where not only quarks~\cite{Claudson:1980yz} but also non-Abelian gauge particles (i.e., gluons)~\cite{Batalin:1976uv,Yildiz:1979vv} can be produced by a chromoelectric field, and to investigate how an imaginary chromomagnetic field modifies the effective Lagrangian and its pole structure.
\item For experiments, an effective imaginary magnetic field can be realized as a Hatano-Nelson lattice, where a lattice with inhomogeneous nonreciprocal hopping on the two-dimensional plane is needed (see Appendix~\ref{sec:Hatano-Nelson}).
By applying an electric field to the third direction in such lattice systems, we can realize the present situation.
\end{itemize}

\section*{Acknowledgments}

A. J. A. is supported by the JAEA Postdoctoral Fellowship Program and partially by the RCNP Collaboration
Research Network Program under Project No. COREnet 057.
This work is supported by the Japan Society for the Promotion of Science (JSPS) KAKENHI under Grant Nos.  JP24K07034 and JP26K17163.
 
\appendix

\section{Standard Schwinger effect}
\label{app:schwinger}

For comparison with the imaginary-magnetic-field case, we briefly review the standard Schwinger effect~\cite{Schwinger:1951nm} in a constant electric field and its modification by a parallel real magnetic field.
 
\subsection{$B=0$ case}

For a constant electric field, corresponding to $B=0$,
the imaginary parts of the one-loop effective Lagrangians
for scalar and Dirac QED are~\cite{Schwinger:1951nm}
\begin{align}
\mathrm{Im}\,\mathcal{L}_{\mathrm{scalar}}
&= \frac{(eE)^2}{16\pi^3} \sum_{n=1}^{\infty} \frac{(-1)^{n-1}}{n^2} \exp\left(-\frac{n\pi m^2}{|eE|}\right), \label{eq:real-Es}
\end{align}
and~\cite{Schwinger:1951nm}
\begin{align}
\mathrm{Im}\,\mathcal{L}_{\mathrm{Dirac}}
&= \frac{(eE)^2}{8\pi^3} \sum_{n=1}^{\infty} \frac{1}{n^2} \exp\left(-\frac{n\pi m^2}{|eE|}\right), \label{eq:real-Ed}
\end{align}
respectively. 
The Dirac result differs from the scalar one by an overall spin factor of $2$ and by the absence of the alternating factor $(-1)^{n-1}$.

\begin{figure}[b]
    \centering
    \includegraphics[width=1\columnwidth]{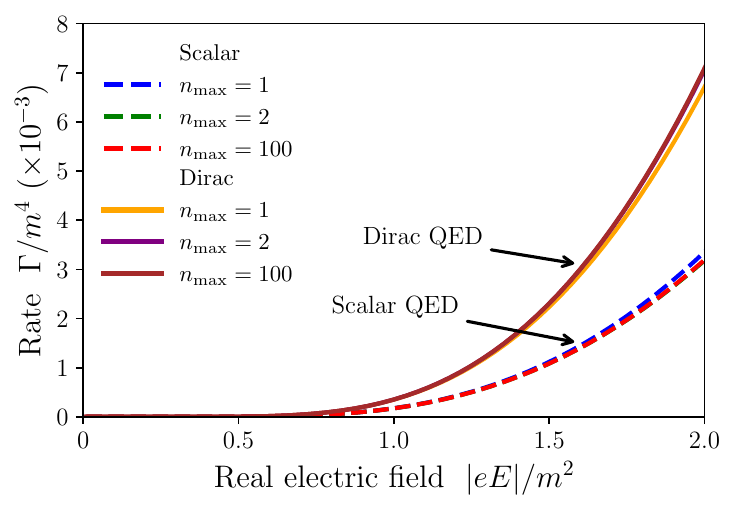} 
    \caption{Vacuum decay rates in a real electric field for scalar and Dirac QED, where we compare the results using the truncation of proper-time poles, $n_{\max}=1,2$ and $100$.
    The Dirac case is larger than the scalar case.
    Both results increase rapidly as the field approaches the critical scale $|eE|\sim m^2$.}
    \label{fig:real-E}
\end{figure}

Figure~\ref{fig:real-E} shows that the vacuum decay rate is exponentially suppressed in the weak-field regime, $|eE|\ll m^2$, by the factor $\exp(-\pi m^2/|eE|)$. 
As the field approaches the characteristic scale, $|eE|\sim m^2$, this exponential suppression becomes weaker, and the vacuum decay rate increases rapidly. 
One can find that the Dirac result is larger than the scalar result due to the spin structure.
In the weak-field regime, it is larger by a factor of two, because the $n=1$ contribution dominates. 
Then, the sums over proper-time poles converge rapidly because higher-$n$ terms are suppressed by
both $1/n^2$ and $\exp(-n\pi m^2/|eE|)$.

\subsection{$B \neq 0$ case}

\begin{figure}[t!]
    \centering
    \includegraphics[width=1\columnwidth]{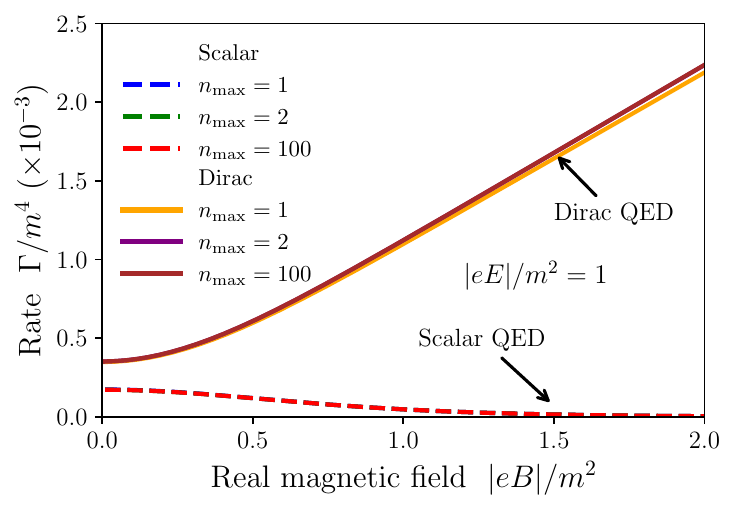}
    \caption{Vacuum decay rates in parallel real electric and
    magnetic fields as a function of $|eB|$.
    A magnetic field suppresses the scalar result, whereas it
    enhances the Dirac result.}
    \label{fig:real-EB}
\end{figure}

A real magnetic field itself does not induce vacuum decay, whereas in the presence of an electric field, it modifies the vacuum decay rate through Landau-level quantization.
The resulting imaginary parts of the one-loop effective Lagrangians are
\begin{align}
\mathrm{Im}\,\mathcal{L}_{\mathrm{scalar}}(B) = \frac{e^2 EB}{16\pi^2} \sum_{n=1}^{\infty} \frac{(-1)^{n-1}}{n} \frac{\exp\left(-\frac{n\pi m^2}{|eE|}\right)}{\sinh\left(n\pi \frac{B}{E}\right)},
\end{align}
and~\cite{Nikishov:1969tt,Bunkin:1969if}
\begin{align}
\mathrm{Im}\,\mathcal{L}_{\mathrm{Dirac}}(B) = \frac{e^2 EB}{8\pi^2} \sum_{n=1}^{\infty} \frac{1}{n} \frac{\exp\left(-\frac{n\pi m^2}{|eE|}\right)}{ \tanh\left(n\pi\frac{B}{E}\right) }.
\end{align}
In the weak magnetic field limit ($B \to 0$), these expressions
smoothly reduce to the pure electric-field formulas, Eqs.~\eqref{eq:real-Es} and \eqref{eq:real-Ed}.
In the strong magnetic field $B/E \gg 1$, the vacuum decay rate in scalar QED is suppressed by $\lim_{x\to\infty}1/\sinh{(x)}=0$, whereas
in Dirac QED, the hyperbolic factor does not suppress it because of $\lim_{x\to\infty}1/\tanh{(x)}=1$, and the vacuum decay rate linearly increases by the overall factor $B$. 
As shown in Fig.~\ref{fig:real-EB}, for a fixed electric field $|eE|/m^2=1$, increasing the magnetic-field strength $|eB|/m^2$ enhances the vacuum decay rate in Dirac QED while suppressing it in scalar QED.

\section{Real Landau levels}
\label{app:landau}

Here, we review the formulas of the Schwinger effect rewritten in terms of Landau-level sums. 

\paragraph{Scalar QED}

We start from the imaginary part of the effective Lagrangian for scalar
QED in real electric and magnetic fields,
\begin{align}
\mathrm{Im}\,\mathcal{L}_{\mathrm{scalar}}(B)
= \frac{e^2 |E| |B|}{16\pi^2}
\sum_{n=1}^{\infty}
\frac{(-1)^{n-1}}{n}
\frac{e^{-\frac{n\pi m^2}{|eE|}}}{\sinh\left(n\pi \frac{|B|}{|E|}\right)}.
\end{align}
To isolate the contributions from individual Landau levels, we use
\begin{align}
\frac{e^{-\frac{n\pi m^2}{|eE|}}}
{\sinh\left(n\pi\frac{|B|}{|E|}\right)} =
2\sum_{\ell=0}^{\infty}
e^{
-\frac{n\pi}{|eE|}
\left(m^2+(2\ell+1)|eB|\right)}.
\end{align}
This naturally defines the effective transverse mass of the $\ell$ th Landau level as $m_{\ell,\mathrm{scalar}}^2 \equiv m^2+(2\ell+1)|eB|.$
Substituting this expansion into the imaginary part gives
\begin{align}
&\mathrm{Im}\,\mathcal{L}_{\mathrm{scalar}}(B) \nonumber\\
&~~=
\frac{e^2 |E||B|}{8\pi^2}
\sum_{n=1}^{\infty}\sum_{\ell=0}^{\infty}
\frac{(-1)^{n-1}}{n}
\left(
e^{-\frac{\pi m_{\ell,\mathrm{scalar}}^2}{|eE|}}
\right)^n .
\end{align}
Exchanging the order of the sums and
using $\sum_{n=1}^{\infty}\frac{(-1)^{n-1}}{n}x^n=\ln(1+x)$ for $|x|<1$,
we finally obtain
\begin{align}
\mathrm{Im}\,\mathcal{L}_{\mathrm{scalar}}(B)
=
\frac{e^2 |E||B|}{8\pi^2}
\sum_{\ell=0}^{\infty}
\ln\left[
1 + e^{
-\frac{\pi m_{\ell,\mathrm{scalar}}^2}{|eE|}
}
\right].
\end{align}

\begin{figure}[t!]
    \centering
    \includegraphics[width=1\columnwidth]{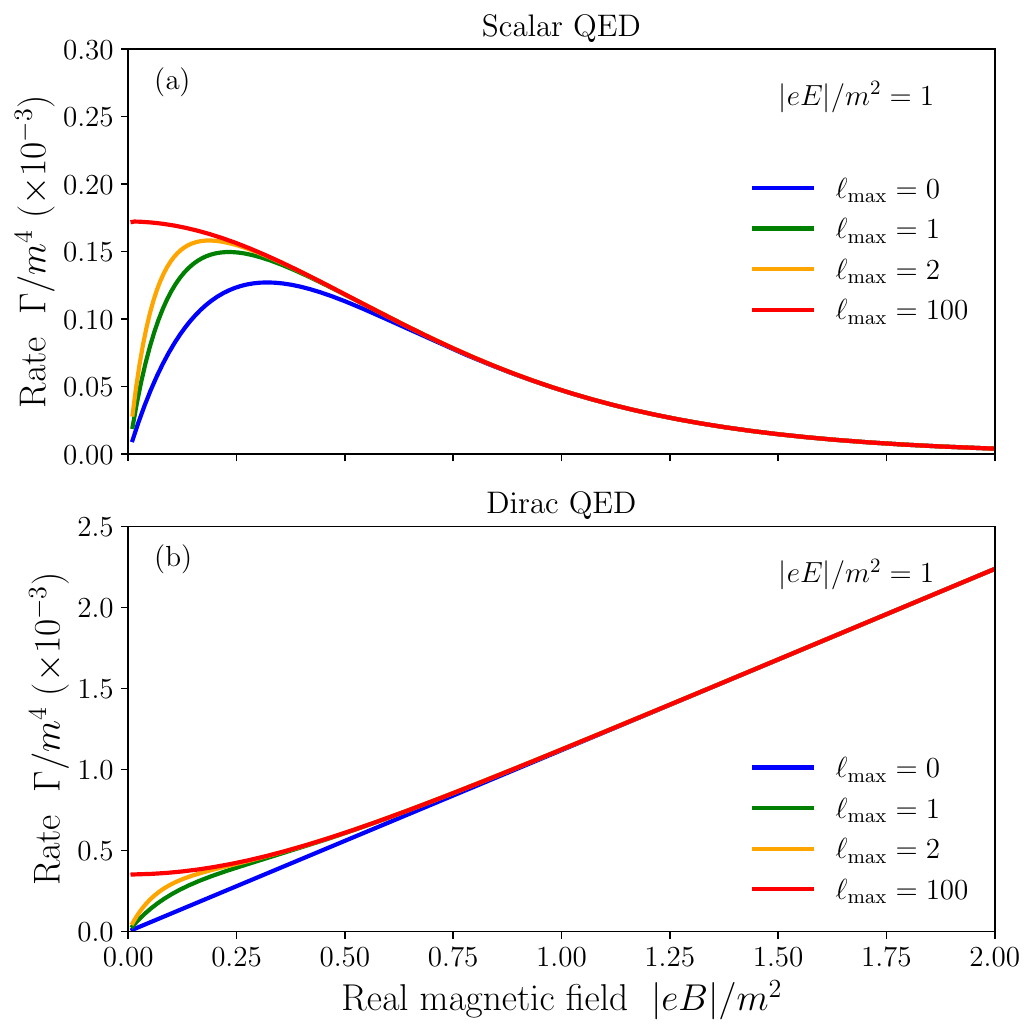}
    \caption{Landau-level decomposition of the vacuum decay rate for (a) scalar and (b) Dirac QED, where we compare the results using truncations of higher Landau levels, $\ell_{\max}=0,1,2,$ and $100$.
    }
    \label{fig:real-EB-landau}
\end{figure}

\paragraph{Dirac QED}

For Dirac QED, we start from the imaginary part~\cite{Nikishov:1969tt,Bunkin:1969if}
\begin{align}
\mathrm{Im}\,\mathcal{L}_{\mathrm{Dirac}}(B)
=
\frac{e^2 |E| |B|}{8\pi^2}
\sum_{n=1}^{\infty}
\frac{1}{n}\frac{e^{-\frac{n\pi m^2}{|eE|}}}{\tanh\left(n\pi\frac{|B|}{|E|}\right)}.
\end{align}
The hyperbolic cotangent can be decomposed as
\begin{align}
\frac{e^{-\frac{n\pi m^2}{|eE|}}}{\tanh\left(n\pi\frac{|B|}{|E|}\right)} =
\sum_{\ell=0}^{\infty}
(2-\delta_{\ell0})
e^{
-\frac{n\pi}{|eE|}
\left(m^2+2\ell |eB|\right)
}.
\end{align}
From this form, we can define the effective transverse mass of the $\ell$ th Landau level as $m_{\ell,\mathrm{Dirac}}^2 \equiv m^2+2\ell|eB|.$
Substituting this expression gives
\begin{align}
&\mathrm{Im}\,\mathcal{L}_{\mathrm{Dirac}}(B) 
\nonumber\\
&~=\frac{e^2 |E||B|}{8\pi^2} \sum_{\ell=0}^{\infty}
(2-\delta_{\ell0}) \sum_{n=1}^{\infty}
\frac{1}{n}
\left(
e^{-\frac{\pi m_{\ell,\mathrm{Dirac}}^2}{|eE|}}
\right)^n .
\end{align}
Using $\sum_{n=1}^{\infty}\frac{x^n}{n}=-\ln(1-x)$ for $|x|<1,$
we obtain
\begin{align}
&\mathrm{Im}\,\mathcal{L}_{\mathrm{Dirac}}(B)\nonumber\\
&~~= -\frac{e^2 |E||B|}{8\pi^2}
\sum_{\ell=0}^{\infty}
(2-\delta_{\ell0})
\ln\left[
1-
e^{
-\frac{\pi m_{\ell,\mathrm{Dirac}}^2}{|eE|}
}
\right].
\end{align}
Equivalently, separating the lowest Landau level (LLL), $\ell=0$, from the higher Landau levels,
\begin{align}
\mathrm{Im}\,\mathcal{L}_{\mathrm{Dirac}}(B)
&=
-\frac{e^2 |E||B|}{8\pi^2}
\Biggl[
\ln\left(
1-e^{-\frac{\pi m^2}{|eE|}}
\right)
\nonumber\\
&\qquad
+2\sum_{\ell=1}^{\infty}
\ln\left(
1-e^{-\frac{\pi m_{\ell,\mathrm{Dirac}}^2}{|eE|}}
\right)
\Biggr].
\end{align}

Figure~\ref{fig:real-EB-landau} shows the Landau-level decomposition of vacuum decay rates at fixed $|eE|/m^2=1$.
At weak magnetic fields, many Landau levels contribute because the spacing between adjacent levels vanishes as $|eB|\to0$.
With increasing $|eB|$, the contributions from higher Landau levels are exponentially suppressed, and the LLL dominates.
For scalar QED [Fig.~\ref{fig:real-EB-landau}(a)], the transverse mass of the LLL increases due to $m_{0,\mathrm{scalar}}^2=m^2+|eB|$, and the vacuum decay rate at large $|eB|$ is suppressed.
In contrast, for Dirac QED [Fig.~\ref{fig:real-EB-landau}(b)], since the LLL does not depend on $B$ due to $m_{0,\mathrm{Dirac}}=m$, its contribution to the rate increases linearly in $|eB|$ through the overall degeneracy factor ${|eB|}/{2\pi}$.

\section{Derivation of complex Landau levels}
\label{app:landau_imag}

Here, we derive Eqs.~\eqref{eq:complex_LL_scalar_rate} and \eqref{eq:complex_LL_dirac_rate} in the complex Landau-level representation of the electric-pole contribution for scalar and Dirac QED in an imaginary magnetic field. 

\paragraph{Scalar QED.}

The electric-pole contribution in scalar QED is
\begin{align}
\mathrm{Im}\,\mathcal{L}_{\mathrm{scalar}}^{(E)}
=
\frac{e^2|E| |B_I|}{16\pi^2}
\sum_{n=1}^{\infty}
\frac{(-1)^{n-1}}{n}
\frac{
\exp\left(-\frac{n\pi m^2}{|eE|}\right)
}{
\sin\left(n\pi\frac{|B_I|}{|E|}\right)
}.
\label{eq:app_scalar_E_poles}
\end{align}
The cosecant factor can be expressed as ${1}/{\sin x} = {2i e^{-ix}}/(1-e^{-2ix}).$
Since the geometric-series expansion of ${1}/{\sin x}$ is not convergent for real $x$, we introduce an Abel regulator $\eta>0$ and write
\begin{align}
\frac{1}{\sin(x-i\eta)}
&=
2i\sum_{\ell=0}^{\infty}
e^{-(2\ell+1)\eta}
e^{-i(2\ell+1)x}.
\label{eq:app_scalar_abel}
\end{align}
For $x\notin\pi\mathbb{Z}$, the original cosecant is
recovered in the limit $\eta\to0^+$.
By substituting this form into the original imaginary part, we obtain
\begin{align}
\mathrm{Im}\mathcal{L}_{\mathrm{scalar}}^{(E)}&=
\lim_{\eta\to0^+}
\mathrm{Re}\Bigg\{
\frac{i e^2|E| |B_I|}{8\pi^2}
\sum_{n=1}^{\infty}
\sum_{\ell=0}^{\infty}
\frac{(-1)^{n-1}}{n}
\nonumber\\
&~~~\times
e^{-(2\ell+1)\eta}
e^{-\frac{n\pi}{|eE|}\left(m^2+i(2\ell+1)|eB_I|\right)}
\Bigg\}.
\label{eq:app_scalar_double_sum}
\end{align}
Here, we take the real part of the regulated expression to obtain a real-valued quantity for finite $\eta>0$.
The factor of $m^2+i(2\ell+1)|eB_I| \equiv M_{\ell,\mathrm{scalar}}^2(B_I)$ corresponds to the complex effective squared mass of the $\ell$ th Landau level for scalar QED.
Using $\sum_{n=1}^{\infty} \frac{(-1)^{n-1}}{n}x^n = \ln(1+x)$ for $|x|<1$, we obtain
\begin{align}
\mathrm{Im}\,\mathcal{L}_{\mathrm{scalar}}^{(E)}
=&~
\lim_{\eta\to0^+}
\mathrm{Re}\Bigg\{
\frac{i e^2|E| |B_I|}{8\pi^2}
\sum_{\ell=0}^{\infty}
e^{-(2\ell+1)\eta}
\nonumber\\
&\quad\times
\ln\left[
1+
\exp\left(
-\frac{\pi M_{\ell,\mathrm{scalar}}^2(B_I)}{|eE|}
\right)
\right]
\Bigg\}.
\label{eq:app_scalar_complex_LL}
\end{align}
This is the Abel-regularized complex Landau-level representation
for the electric-pole contribution in Eq.~\eqref{eq:complex_LL_scalar_rate}.

\paragraph{Dirac QED.}

For Dirac QED, the electric-pole contribution is
\begin{align}
\mathrm{Im}\,\mathcal{L}_{\mathrm{Dirac}}^{(E)}
=&~
\frac{e^2|E| |B_I|}{8\pi^2}
\sum_{n=1}^{\infty}
\frac{1}{n}
\frac{\exp\left(-\frac{n\pi m^2}{|eE|}\right)}{
\tan\left(n\pi\frac{|B_I|}{|E|}\right)}.
\label{eq:app_dirac_E_poles}
\end{align}
The cotangent factor $1/\tan x = i(1+e^{-2ix})/(1-e^{-2ix})=i[1+2e^{-2ix}/(1-e^{-2ix})]$ is written as, using an Abel regulator $\eta>0$, 
\begin{align}
\frac{1}{\tan(x-i\eta)}
&=
i\sum_{\ell=0}^{\infty}
g_\ell e^{-2\ell\eta}e^{-2i\ell x},
\label{eq:app_dirac_abel}
\end{align}
where we introduced the degeneracy factor $g_\ell=2-\delta_{\ell0}$:
$g_0=1$ for the lowest Landau level, while $g_{\ell}=2$ for the higher Landau levels.

Substituting Eq.~\eqref{eq:app_dirac_abel} into
Eq.~\eqref{eq:app_dirac_E_poles} gives
\begin{align}
\mathrm{Im}\,\mathcal{L}_{\mathrm{Dirac}}^{(E)}
=&~
\lim_{\eta\to0^+}
\mathrm{Re}\Bigg\{
\frac{i e^2|E| |B_I|}{8\pi^2}
\sum_{n=1}^{\infty}
\sum_{\ell=0}^{\infty}
\frac{g_\ell}{n}
e^{-2\ell\eta}
\nonumber\\
&\quad\times
\exp\left[
-\frac{n\pi}{|eE|}
\left(m^2+2i\ell |eB_I|\right)
\right]
\Bigg\}.
\label{eq:app_dirac_double_sum}
\end{align}
The corresponding complex effective squared masses are $ m^2+2i\ell |eB_I| \equiv M_{\ell,\mathrm{Dirac}}^2(B_I)$.
Using $\sum_{n=1}^{\infty}\frac{x^n}{n} = -\ln(1-x)$ for $|x|<1$,
we obtain
\begin{align}
\mathrm{Im}\,\mathcal{L}_{\mathrm{Dirac}}^{(E)}
=&~
\lim_{\eta\to0^+}
\mathrm{Re}\Bigg\{
-\frac{i e^2|E| |B_I|}{8\pi^2}
\sum_{\ell=0}^{\infty}
g_\ell e^{-2\ell\eta}
\nonumber\\
&\quad\times
\ln\left[
1-
\exp\left(
-\frac{\pi M_{\ell,\mathrm{Dirac}}^2(B_I)}{|eE|}
\right)
\right]
\Bigg\}.
\label{eq:app_dirac_complex_LL}
\end{align}
This is Eq.~\eqref{eq:complex_LL_dirac_rate}.

\section{Inhomogeneous Hatano-Nelson lattice} 
\label{sec:Hatano-Nelson}

An imaginary magnetic field can be realized theoretically in lattice systems (and also experimentally, if the corresponding Hamiltonian can be implemented).
In this Appendix, as an example of such a realization, we consider a two-dimensional lattice system with a nonreciprocal interaction based on the Hatano-Nelson model~\cite{Hatano:1996ntg,Hatano:1997}, described by the Hamiltonian

\begin{equation}
H=\sum_{\mathbf r}\sum_{\mu=x,y}
\left[
t_\mu^{\mathrm f}(\mathbf r)c_{\mathbf r+a\hat\mu}^{\dagger}c_{\mathbf r}
+t_\mu^{\mathrm b}(\mathbf r)c_{\mathbf r}^{\dagger}c_{\mathbf r+a\hat\mu}
\right],
\end{equation}
where $c_{\mathbf r}^{\dagger}$ and $c_{\mathbf r}$ are the creation and annihilation operators at lattice site $\mathbf r=(x,y)$, respectively, $a$ is the lattice spacing, and $\hat\mu$ is the unit vector along the $\mu$ direction. $t_\mu^{\mathrm f}$ and $t_\mu^{\mathrm b}$ are the forward and backward hopping parameters, respectively.
In reciprocal systems, $t_\mu^{\mathrm f} = t_\mu^{\mathrm b}$.

\begin{figure}[b!]
    \centering    
    \includegraphics[width=1\columnwidth]{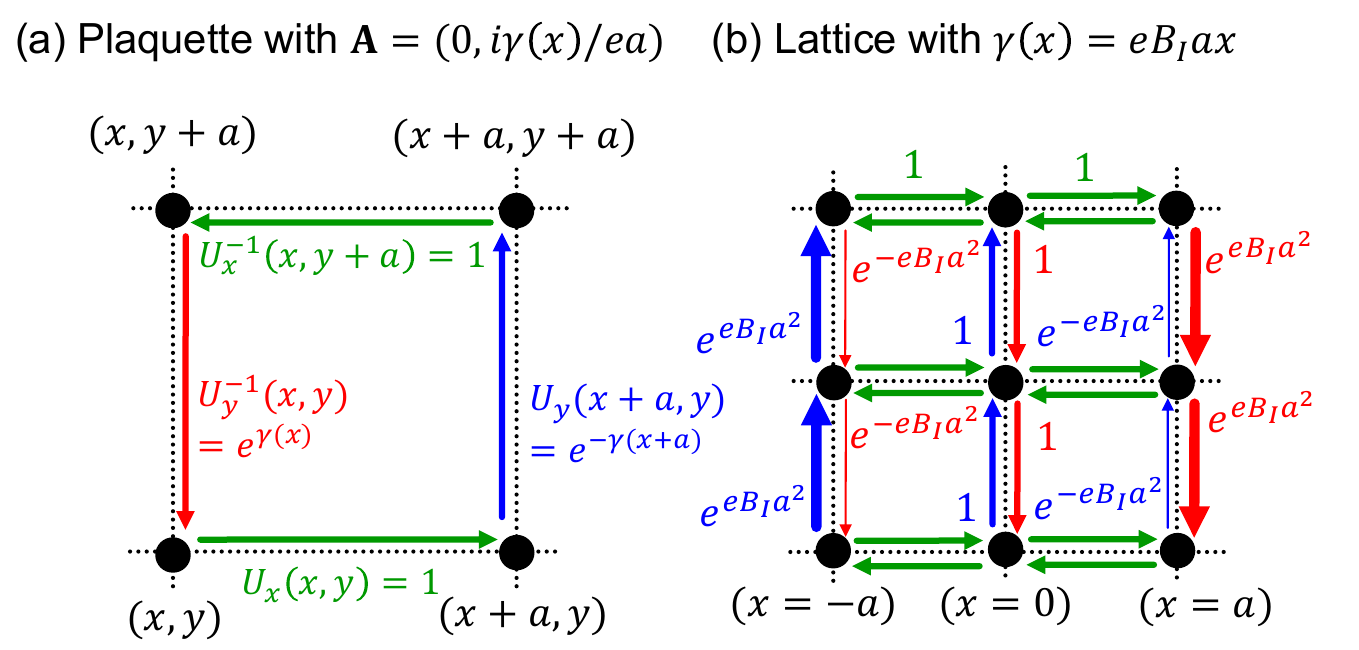}
    \caption{Imaginary magnetic field realized on a two-dimensional lattice with inhomogeneous (i.e., $x$-dependent) nonreciprocal parameters $\gamma(x)$.
    (a) Plaquette in the imaginary gauge potential $\mathbf{A}=(0,A_y(x))$ with $A_y=\frac{i\gamma(x)}{ea}$.
    (b) Lattice with $\gamma(x)=eB_I ax$.}
    \label{fig:HatanoNelson}
\end{figure}

Here, we consider a spatially dependent nonreciprocity
\begin{align}
t_x^{\mathrm f}(\mathbf r)&=t_x^{\mathrm b}(\mathbf r)=t,\\
t_y^{\mathrm f}(\mathbf r)&=t e^{-\gamma(x)},\
t_y^{\mathrm b}(\mathbf r)=t e^{+\gamma(x)}, \label{eq:ty}
\end{align}
where $\gamma(x)$ characterizes the nonreciprocity along the $y$ direction.
In the conventional Hatano-Nelson case, $\gamma$ is independent of $x$, whereas $\gamma$ depends on $x$ in an {\it inhomogeneous Hatano-Nelson} case.

The link variable is defined as
\begin{equation}
U_\mu(\mathbf{r})=
\exp\left[
ie\int_{\mathbf{r}}^{\mathbf{r}+a\hat{\mu}}
\mathbf{A}(\boldsymbol{\ell})\cdot d\boldsymbol{\ell}
\right],
\end{equation}
where $e$ is the particle charge, $\mathbf{A}=(A_x,A_y)$ is the gauge potential, $a$ is the lattice spacing, and $\boldsymbol{\ell}$ denotes a point along the integration path.

We choose the Landau gauge, $\mathbf{A}=(0,A_y(x))$, with the $x$-dependent imaginary gauge potential
\begin{equation}
A_y(x)=\frac{i\gamma(x)}{ea}.
\end{equation}
Then, the link variables become
\begin{equation}
U_x(\mathbf{r})=1,\qquad
U_y(\mathbf{r})=e^{ieaA_y(x)}=e^{-\gamma(x)}.
\end{equation}
Thus, the forward and backward link variables, $U_y(x)$ and $U_y^{-1}(x)$, correspond to the $x$-dependent nonreciprocal hopping parameters in Eq.~\eqref{eq:ty}.

The plaquette variable is defined as the product of link variables along the four links:
\begin{align}
P_{xy}(x)
&=U_x(x,y)U_y(x+a,y)
U_x(x,y+a)^{-1}U_y(x,y)^{-1} \nonumber\\
&=\frac{U_y(x+a)}{U_y(x)} \nonumber\\
&=e^{-[\gamma(x+a)-\gamma(x)]},
\end{align}
as shown in Fig.~\ref{fig:HatanoNelson}(a).
When we substitute a linear profile $\gamma(x)=eB_I ax$, we can find that the plaquette variable is spatially uniform:
\begin{equation}
P_{xy}(x)=e^{-eB_Ia^2},
\end{equation}
Thus, although the profile of $\gamma(x)$ is inhomogeneous, each plaquette variable is homogeneous, and the corresponding magnetic field is also homogeneous, as in Fig.~\ref{fig:HatanoNelson}(b).

We note that the extension to a three-dimensional lattice is straightforward.
In addition, if one imposes periodic boundary conditions on this system, the $x$-direction link variables at the lattice edges in the $y$ direction must be modified to ensure the homogeneity of plaquette variables.

\bibliography{references}

\end{document}